\documentclass[review]{elsarticle}

\usepackage[a4paper, total={6in, 9in}]{geometry}

\journal{International Journal of Heat and Mass Transfer}

\begin{document}

\begin{frontmatter}

\title{A systematic study of turbulent heat transfer over rough walls}

%

\author[kit]{Pourya Forooghi\corref{mycorrespondingauthor}}
\author[hs]{Matthias Stripf}
\author[kit]{Bettina Frohnapfel}


\address[kit]{Institut f\"ur Str\"omungsmechanik, Karlsruher Institut f\"ur Technologie, 76131 Karlsruhe, Germany}
\address[hs]{Forschungsbereich Thermofluiddynamik und effiziente Energiewandlung, Hochschule Karlsruhe, 76133 Karlsruhe, Germany}

\begin{abstract}
Direct Numerical Simulations are used to solve turbulent flow and heat transfer over a variety of rough walls in a channel. The wall geometries are exactly resolved in the simulations. The aim is to understand the effect of roughness morphology and its scaling on the  augmentation of heat transfer relative to that of skin friction. A number of realistic rough surface maps obtained from the scanning of gas turbine blades and internal combustion engines as well as several artificially generated rough surfaces are examined. In the first part of the paper, effects of statistical surface properties, namely surface slope and roughness density, at constant roughness height are systematically investigated, and it is shown that Reynolds analogy factor (two times Stanton number divided by skin friction coefficient) varies meaningfully but moderately with the surface parameters except for the case with extremely low slope or density where the Reynolds analogy factor grows significantly and tends to that of a smooth wall. In the second part of the paper, the roughness height is varied (independently in both inner and outer units) while the geometrical similarity is maintained. Considering all the simulated cases, it is concluded that Reynolds analogy factor correlates fairly well with the equivalent sand roughness scaled in inner units and asymptotically tends to a plateau.
\end{abstract}


\end{frontmatter}

\section{\label{sec:intro}Introduction}
Convection heat transfer over rough walls find applications in several areas of industry. Gas turbine blades are often rough on account of coating or degradation during service \cite{Bons01}. Novel additive manufacturing techniques, which can be used for devices such as heat exchangers, create very rough surfaces \cite{Stimpson16}. In-cylinder heat transfer in IC-engines \cite{Weidenlener17} and ice accretion on aircrafts \cite{Mcclain17} are other examples of heat transfer over rough surfaces.

Roughness can significantly alter both skin friction and heat transfer on the wall. A large portion of the published research in this field is devoted to the calculation of equivalent sand roughness $k_s$ for different roughness types and to the understanding of the relation between this quantity and the geometry of roughness. Equivalent sand-grain roughness, first defined by Schlichting in his pioneering work on roughness \cite{Schlichting36}, is the size of sand-grain out of Nikuradse's experiments \cite{Nikuradse33} that produces the same skin friction coefficient as the arbitrary surface of interest\footnote{In the fully-rough regime.}. Schlichting was the first who experimentally determined the equivalent sand-grain roughness for several artificially roughened surfaces \cite{Schlichting36}. Finding a correlation between $k_s$ and surface geometrical properties is particularly difficult for naturally roughened surfaces thanks to their disparate and stochastic nature. Several of the existing correlations are reviewed by Flack and Schultz \cite{FlackSchultz10}. Recently, Direct Numerical Simulation (DNS) is also utilized to produce systematic data required for the development of a universal $k_s$ correlation \cite{Forooghi17a,Thakkar17}.

Equivalent sand-grain roughness $k_s$ is predominantly used in the engineering CFD tools to represent the effect of roughness. It is well established that the main effect of roughness on the mean velocity profile is  a downward shift in the logarithmic region, which may be parametrized by $k_s$ \cite{Jimenez04}. In turbulence modeling this can be a basis for the modification of wall functions. It is also widely accepted that, except for a region in the immediate vicinity of the rough wall, the structure of turbulence is unaffected by the roughness \cite{Flack07,Squire16}. In view of the above facts, the equivalent sand-grain roughness approach is expected to function for the flow over rough walls as well as for the flow over smooth walls, as long as the right $k_s$ value is known a priori. For the prediction of heat transfer, however, relying solely on $k_s$ can lead to error as the increase in heat transfer due to roughness is not proportional to that of momentum transfer. In other words, Reynolds analogy does not hold for rough surfaces \cite{DippreySabersky63, Bons05, Aupoix15}.

To address the above-mentioned shortcoming, Aupoix \citep{Aupoix15} suggests a modified $k_s$-based approach in the framework of Reynolds-averaged Navier Stokes (RANS) modeling, in which a corrected expression for turbulent Prantdl number above rough walls is utilized, thereby heat transfer predictions are meaningfully improved. The likes of this modified approach, however, require experimental or high fidelity numerical data for calibration and validation. Aupoix \cite{Aupoix15} use a Discrete Element Method (DEM) for generating a database, based on which the model coefficients can be tuned. In DEM, roughness is effectively `modeled' using source terms in the momentum, energy and turbulence transport equations (see \cite{Taylor85, Stripf08} for examples of DEM in modeling of fully turbulent and transitional flows, respectively). While being computationally less costly, DEM does not provide a level of fidelity comparable to full-geometry resolving DNS (simply referred to as DNS hereinafter). Therefore, as pointed out by Aupoix, further progress in the prediction of heat transfer over rough walls can be achieved by refining the available models through two possible routes: use of available/new experimental databases or creation of a DNS database. The present paper follows the second route.

A number of experimental reports on turbulent heat transfer over rough surfaces are available in literature. Ligrani et al. \cite{Ligrani79} report measurements of skin friction coefficient and Stanton number for developing boundary layers over a plate roughened by packed spheres. Stimpson et al. \cite{Stimpson16} measured pressure drop coefficient and Nusselt number over additively manufactured surfaces in ducts with different hydraulic diameters.  Stripf et al. \cite{Stripf05} measure heat transfer on a high-pressure turbine vane roughened by distributed truncated cones, and studied the effect of roughness density and upstream turbulence level. In their comprehensive experimental campaign, Bons \cite{Bons05,Bons02} and Bons and McClain \cite{BonsMcclain04} examine a number of realistic roughness geometries from gas turbine blades -- each geometry representative of a surface degradation mechanism. These authors also systematically study the effects of pressure gradient and upstream turbulence level on both heat transfer and skin friction. Bons \cite{Bons05} uses the Reynolds analogy factor
\begin{equation}
RA=\frac{2St}{C_f}
\label{eq:RA}.
\end{equation}
of a rough wall normalized by that of the corresponding smooth wall $RA_0$ to quantify the relative augmentation of heat and momentum transfer due to roughness. This concept is used extensively in the present paper.

DNS provides not only accurate results but also absolute access to all flow variables everywhere in the computational domain, making it an ideal choice for generating benchmark data. Only a few DNS reports can be found in the open literature in which heat transfer is taken into account \cite{Miyake01,Nagano04, Leonardi15,Orlandi16}, which mainly focus on simple 2D geometries. Among the above references, \cite{Orlandi16} studies one 3D roughness geometry generated by identical cubes. The limited amount of existing DNS data for heat transfer over 3D roughness calls for further work in this area, and the present paper attempts to partially fill this gap.

The present paper provides DNS results for skin friction and heat transfer over several 3D rough surfaces, including both artificial and realistic roughness samples. We systematically study the effect of roughness morphology at fixed roughness height and that of roughness height (in inner and outer scales) at fixed roughness morphology \footnote{In the present paper, `morphology' is an umbrella term referring to the statistical surface properties, which can be changed independent of the characteristic height of roughness (or simply height of roughness elements when the surface is roughened by such elements). ``A change of roughness height at fixed morphology'' means that the geometrical similarity of roughness is maintained while the characteristic roughness height is scaled up or down in inner/outer units.}. In total, 25 simulations at friction Reynolds number of 500 are run in a fully-developed channel flow configuration. The results are expected to provide a basis for the calibration and evaluation of the models used by engineers. Apart from that, such a systematic study can shed light on the physics of heat transfer augmentation over rough walls. Two important questions that the present results are sought to answer are the following. (1) What are the relevant scales in determining the Reynolds analogy factor over rough walls? (2) Whether and to what extent is this factor affected by merely a change in the surface morphology?

The paper is organized in the following way. Section 2 is devoted to the introduction of the roughness samples -- both artificial and realistic. Numerical solution is explained in section 3. In section 4, the simulation results are presented. In the present paper, only the integral quantities of direct engineering interest, e.g $C_f$, $St$ and $RA/RA_0$ are discussed. Finally, the main findings are summarized in section 5.

\section{\label{sec:samples}Roughness samples}
Two types of surface roughness samples are studied in the present paper. (1) Artificial roughness generated by the distribution of roughness elements on a reference smooth plane (bottom plane). (2) Realistic roughness based on scanning of industrial rough surfaces. In the following each type is explained.
\subsection{\label{sec:samples1}Artificial roughness}
Generation of roughness by distributing roughness elements on a smooth surface is common in both experimental and numerical communities. Such roughness geometries are straightforward to parametrize, thus provide the possibility to isolate the effects of different parameters. Study of roughness morphology in this paper is mainly based on the artificial samples.

A full description of our roughness generation approach and the shapes of elements is available in \cite{Forooghi17a}. In what follows, only the aspects important for the present study are discussed for brevity. To form each roughness sample, a certain number of elements -- calculated from a prescribed total frontal area --  are generated. The height of each roughness element $k$ follows a  random function with normal distribution, prescribed mean $k_m$ and standard deviation $\sigma_k$. The positions of the roughness elements on the bottom plane are also determined randomly with a uniform distribution. Although such an artificial roughness still lacks the complexity of a realistic roughness, our attempt is to mimic some features of realistic roughness by randomness in size and positioning of elements. Therefore, the presently studied artificial samples can be considered a closer representation of realistic roughness than those generated by same-size regularly distributed elements, widely used in the literature.

\begin{figure}[h!]
	\begin{center}
		\includegraphics*[trim={0.5cm 0cm 0.8cm 0cm},clip,width=0.6\textwidth]{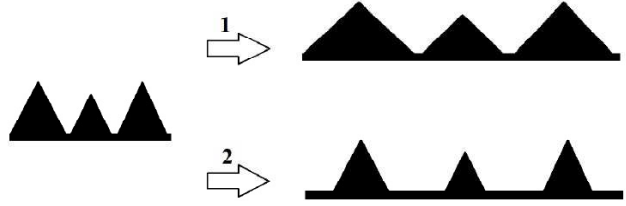}
		\caption{\label{fig:elements} Side views of two types of roughness elements used in the paper: type A (left) and type B (right). $k$ and $D$ denote height and effective diameter of the element (the diameter of a cylinder with the same frontal area), respectively.}
	\end{center}
\end{figure}

Two shapes of roughness elements are used in the present study as shown in figure \ref{fig:elements}. We label the elements on the left and right hand side of this figure A and B, respectively. The roughness elements are axisymmetric. Element A has a less steep side profile and roughly resembles a slightly truncated cone. Element B has a very steep side but is flat on top, hence is more similar to a cylinder with rounded top rim. The effective diameter of the element $D$ is calculated by dividing the frontal projected area of the element by its height $k$. With the same height $k$, A and B have the same frontal area, hence the same $D$ (except for the \textit{stretched} samples explained below). The arithmetic mean height and effective diameter of the elements are denoted by $k_m$ and $D_m$. We also define an `effective mean element spacing' $S_m$ for a rough surface, which is the diameter of a circle with $A_m$ area, when $A_m$ is the area occupied by one roughness element, i.e. total bottom plane area divided by the number of elements.

\begin{figure}[h!]
	\begin{center}
		\includegraphics*[trim={0cm 0cm 0cm 0cm},clip,width=0.6\textwidth]{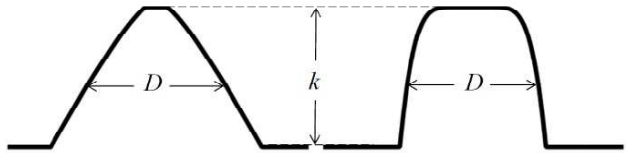}
		\caption{\label{fig:schematic} Schematic representation of how the slope (top arrow) or density (bottom arrow) of a certain roughness geometry (the one on the left hand side) varies.}
	\end{center}
\end{figure}

In this paper we are mainly interested in two morphological properties: `density' and `slope'. Figure \ref{fig:schematic} illustrates how a variation in density differs from a variation in slope. When the slope of a roughness changes it means that the geometry is laterally scaled up or down but the elevation of the rough surface remains the same (arrow number 1 in the figure shows a decrease in slope). In other words the aspect ratio of the features in the roughness geometry changes. When the density of roughness changes, on the other hand, it means that the shape of roughness elements are preserved but they are distributed with larger or smaller spacings, in other words more sparsely or densely (arrow number 2 in the figure shows a decrease in density). Based on this explanation, the `slope' of the presently studied artificial roughness can be measures using the $k_m/D_m$ ratio (high $k_m/D_m$ indicates steepness). Similarly, the `density' can be measured by the $D_m/S_m$ ratio (high $D_m/S_m$ indicates high density).

\begin{figure}[h!]
	\begin{center}
		\includegraphics*[trim={0.5cm 1.5cm 0.8cm 1.5cm},clip,width=0.32\textwidth]{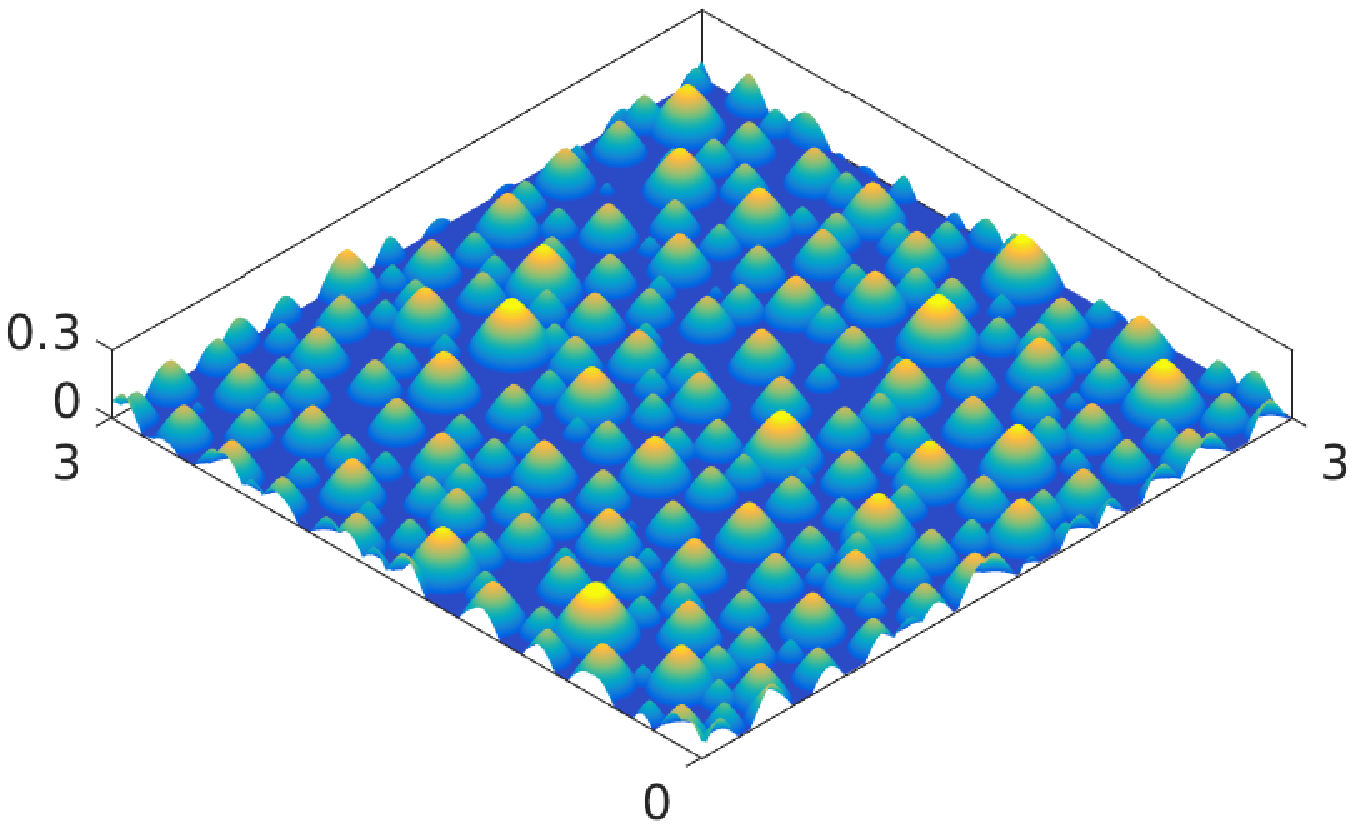}
		\includegraphics*[trim={0.5cm 1.5cm 0.8cm 1.5cm},clip,width=0.32\textwidth]{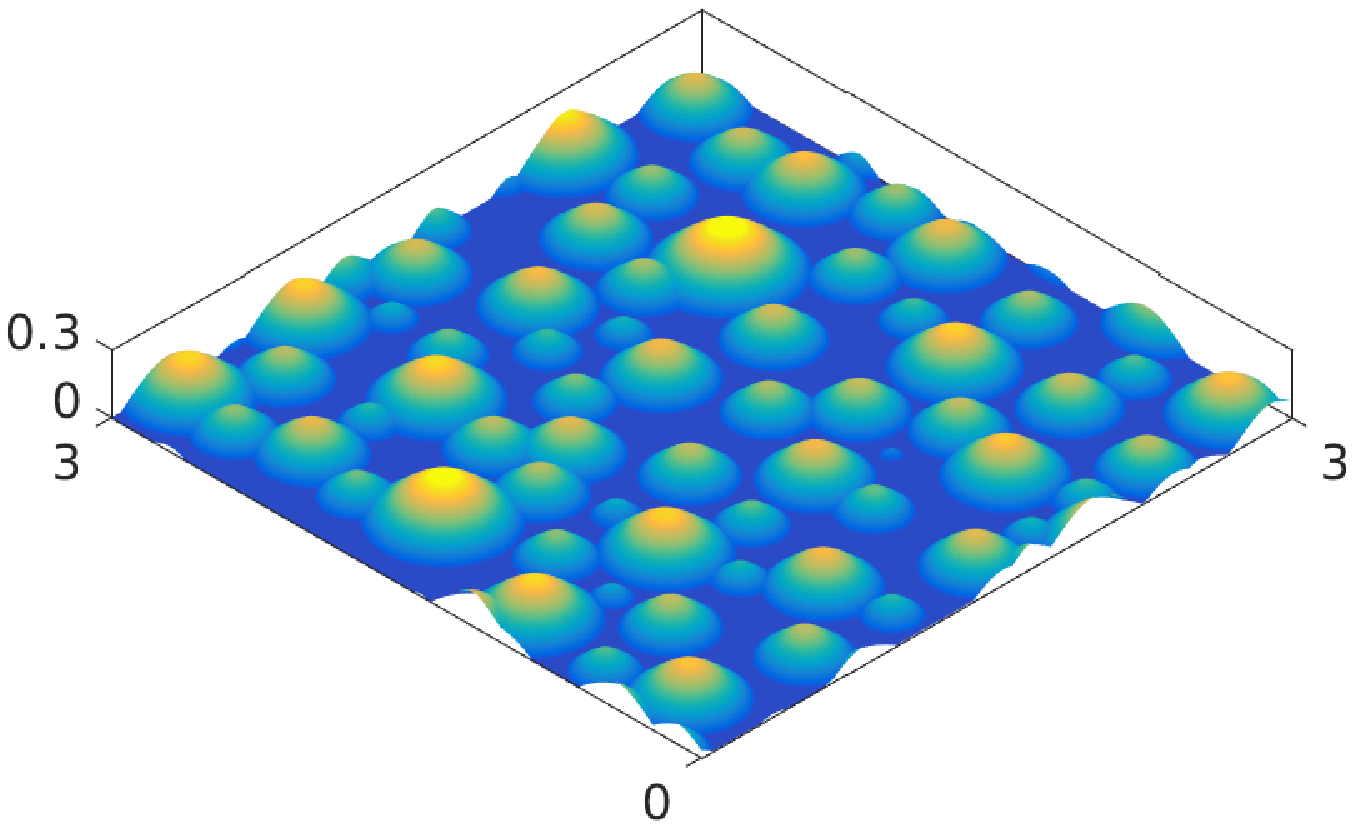}
		\includegraphics*[trim={0.5cm 1.5cm 0.8cm 1.5cm},clip,width=0.32\textwidth]{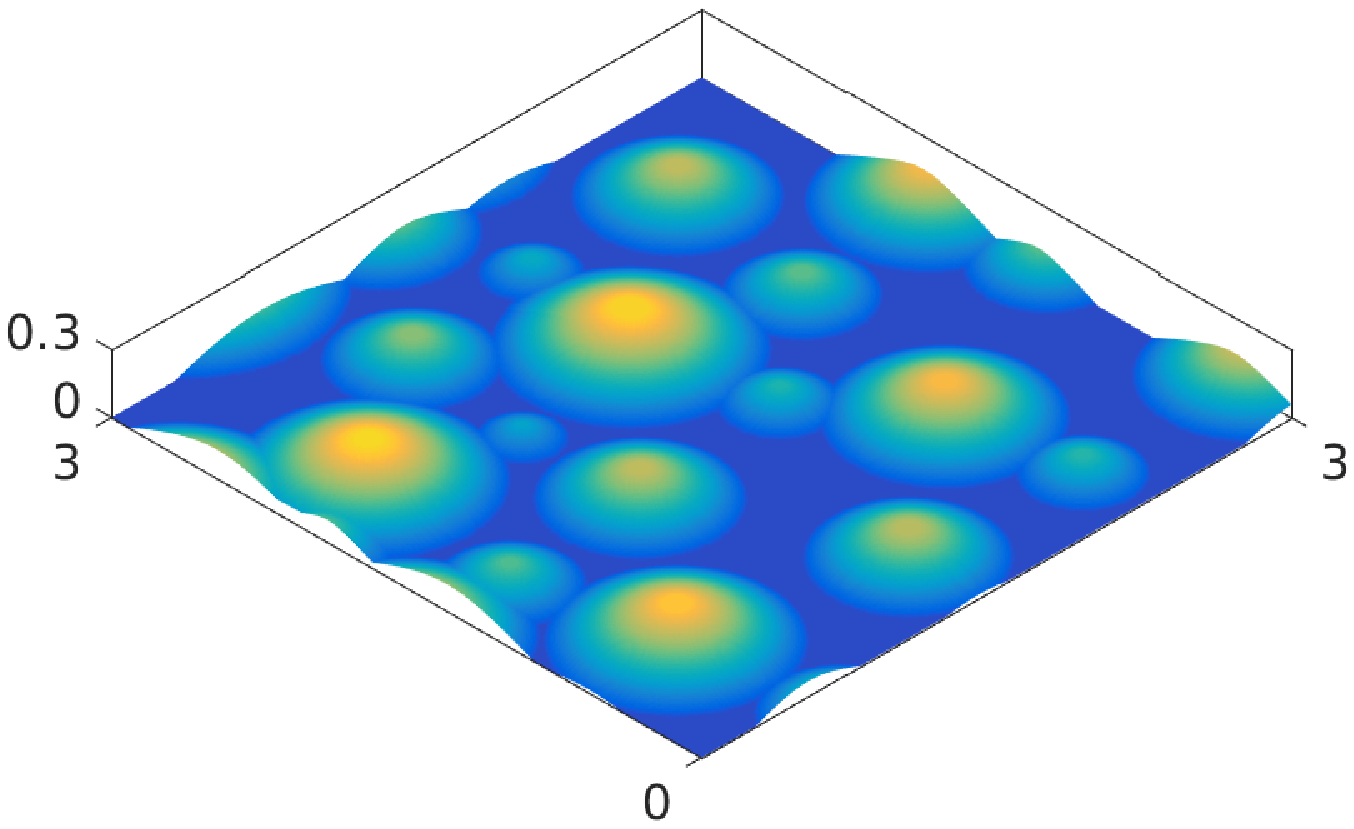}
		\includegraphics*[trim={0.5cm 1.5cm 0.8cm 1.5cm},clip,width=0.32\textwidth]{AA80}
		\includegraphics*[trim={0.5cm 1.5cm 0.8cm 1.5cm},clip,width=0.32\textwidth]{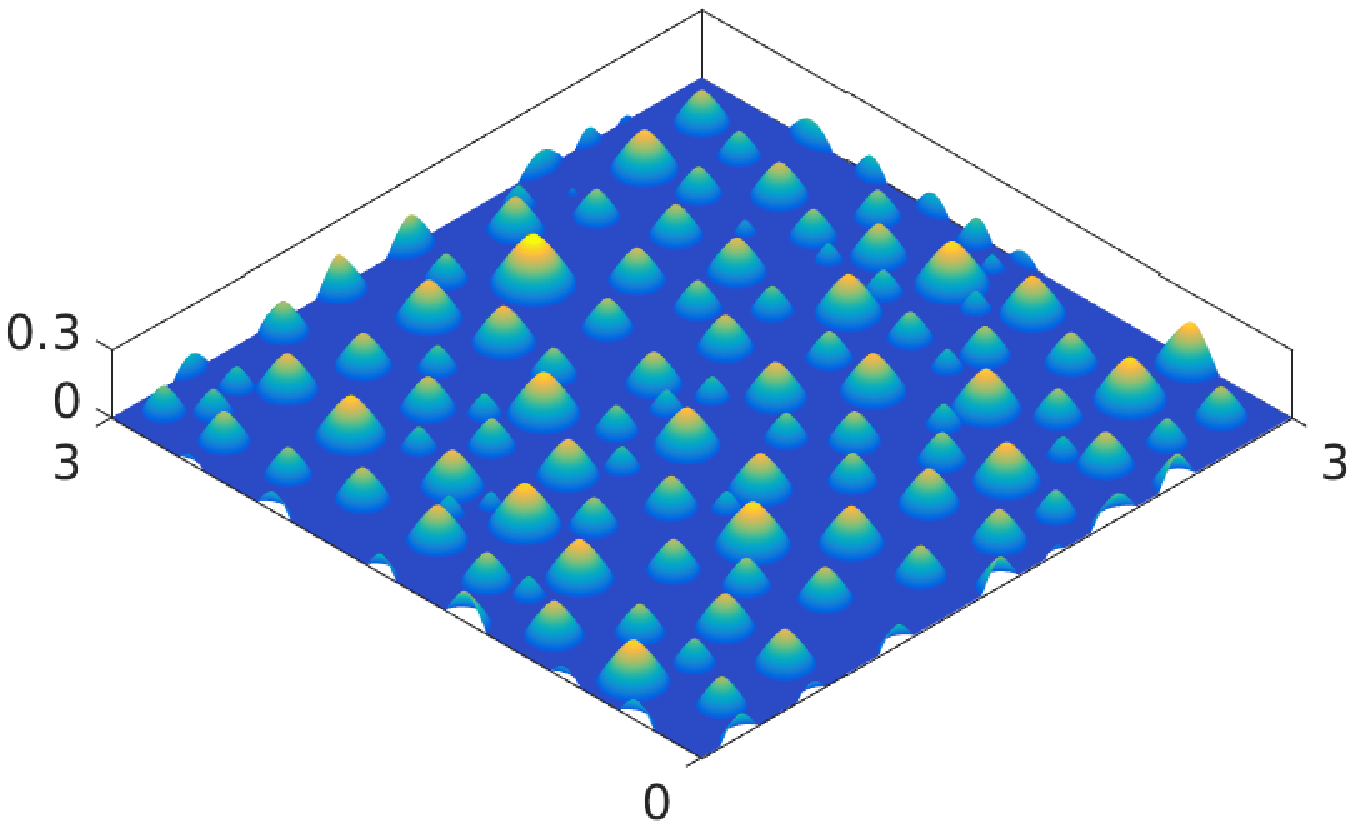}
		\includegraphics*[trim={0.5cm 1.5cm 0.8cm 1.5cm},clip,width=0.32\textwidth]{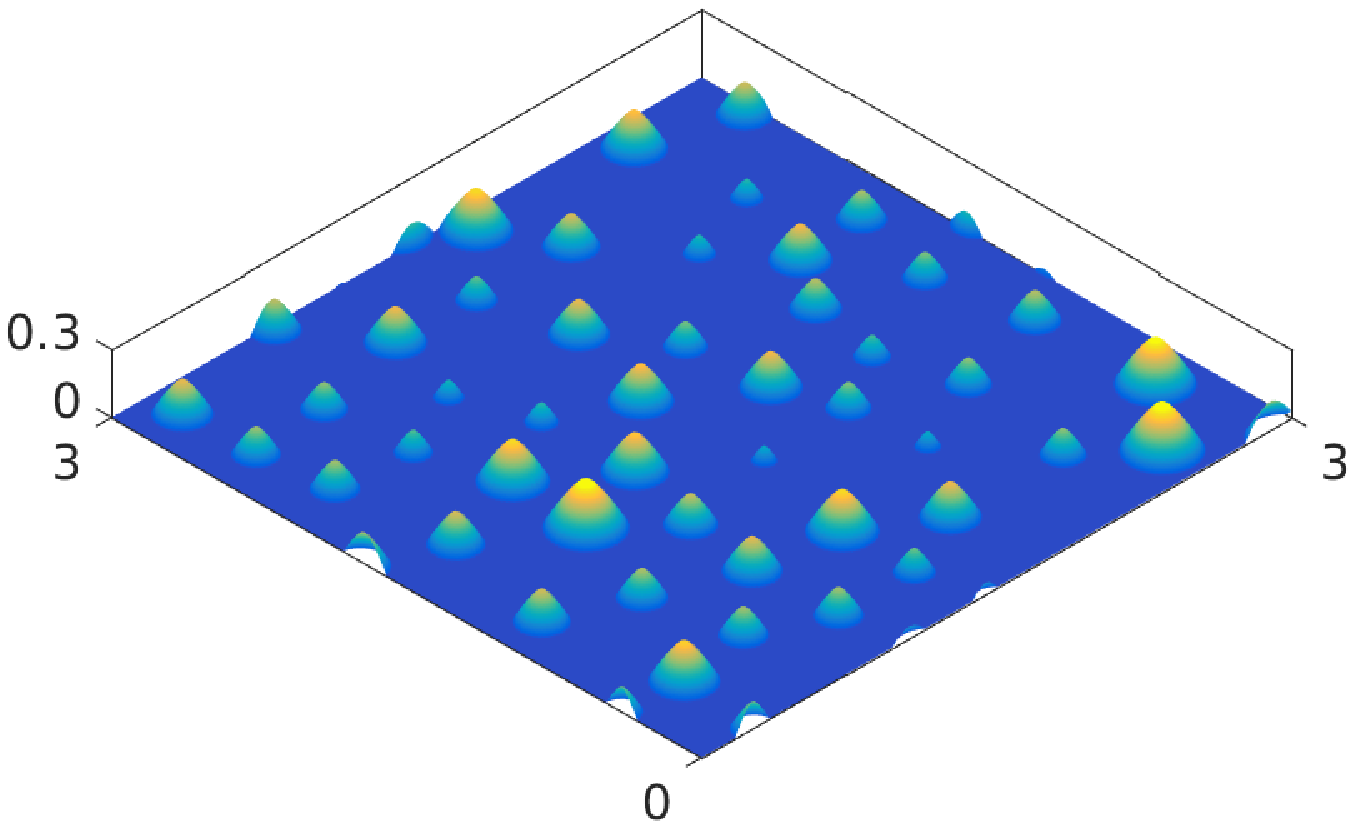}
		\includegraphics*[trim={0.5cm 1.5cm 0.8cm 1.5cm},clip,width=0.32\textwidth]{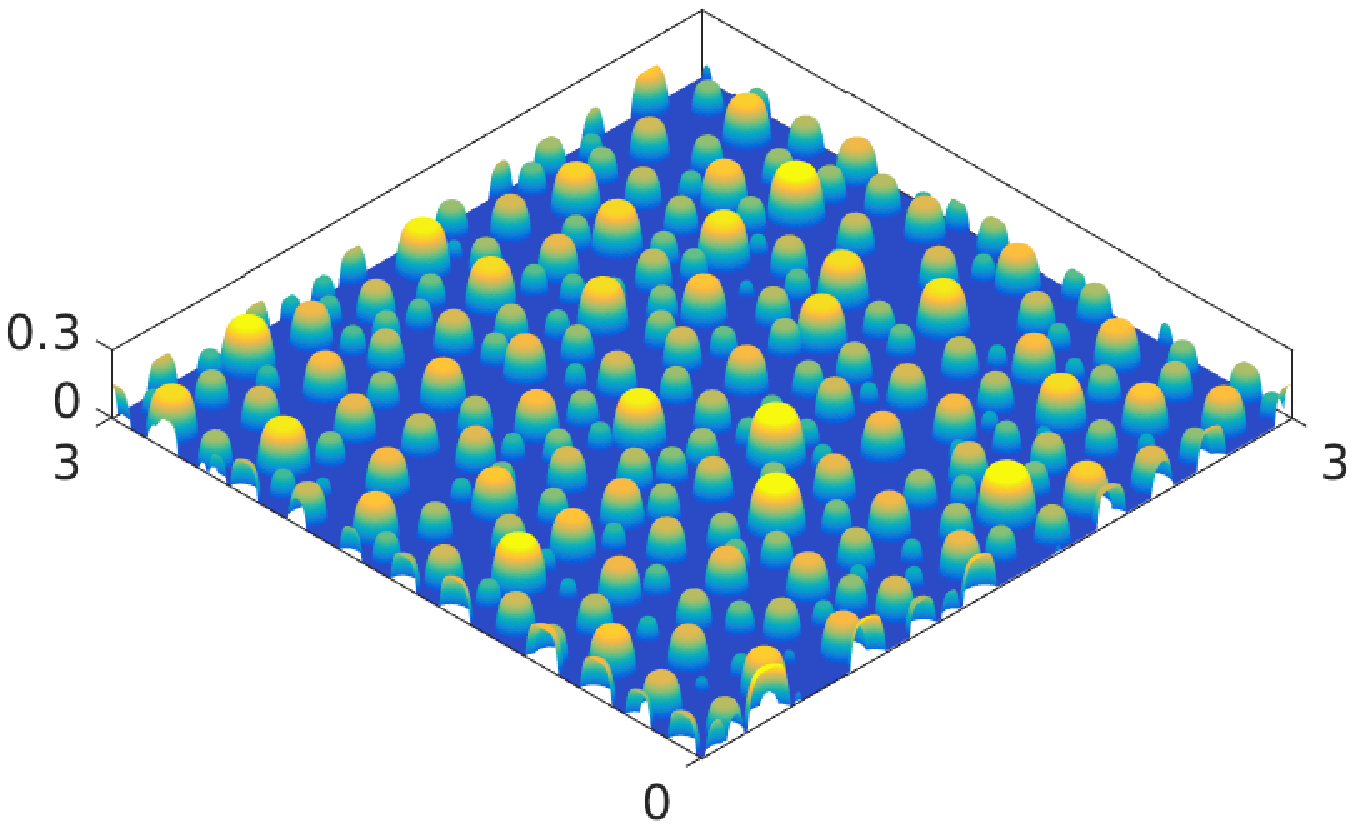}
		\includegraphics*[trim={0.5cm 1.5cm 0.8cm 1.5cm},clip,width=0.32\textwidth]{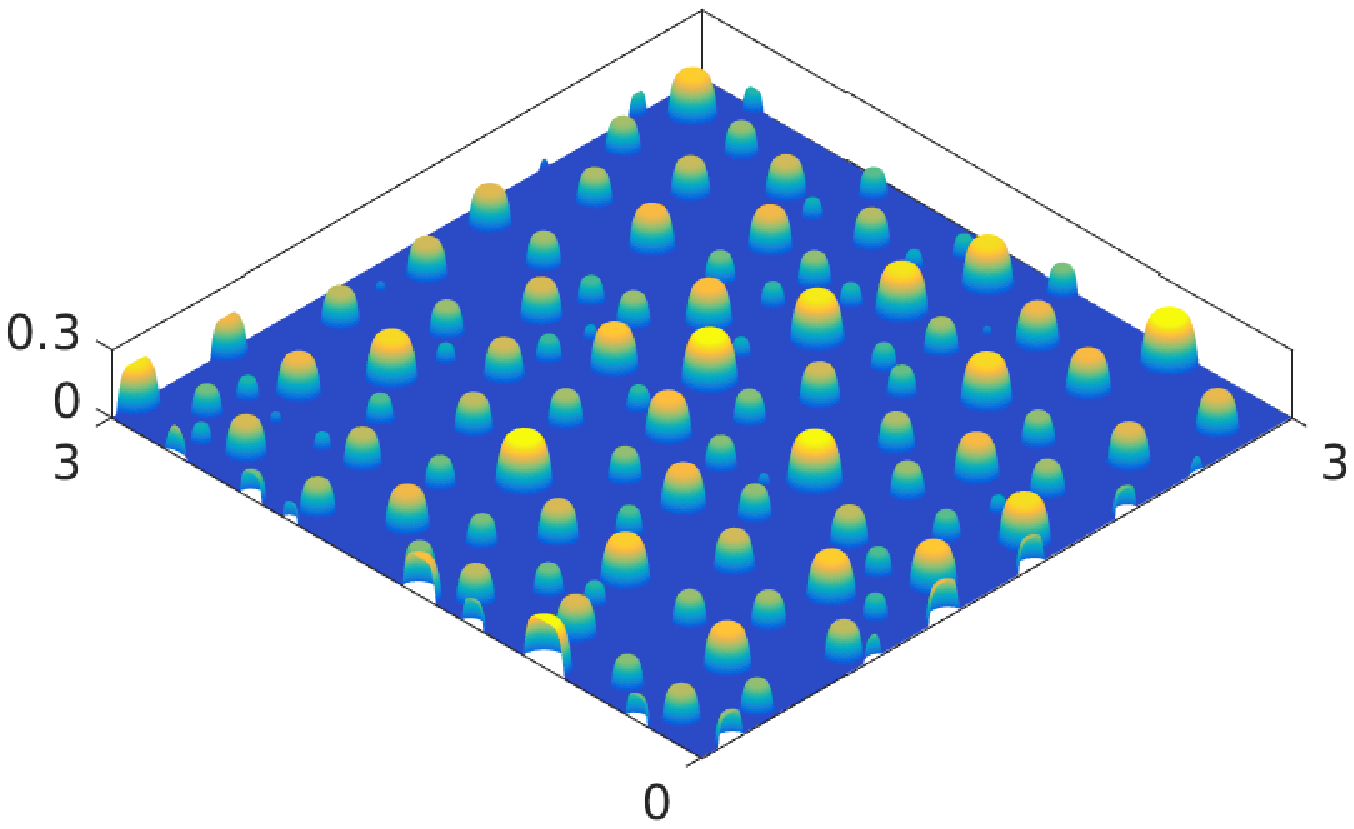}
		\includegraphics*[trim={0.5cm 1.5cm 0.8cm 1.5cm},clip,width=0.32\textwidth]{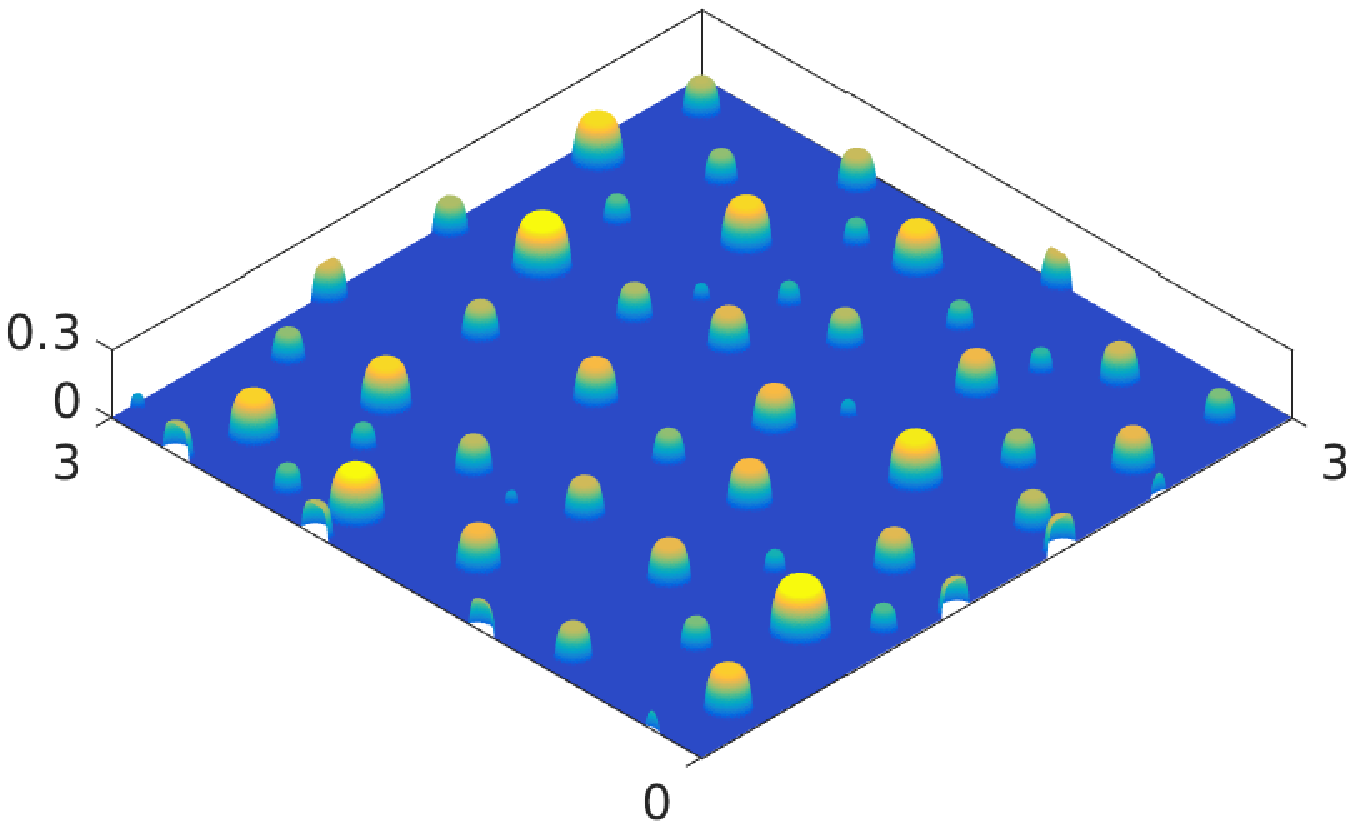}
		\caption{\label{fig:samples} Patches from different roughness samples. Top row: A-s set (from left to right: A-s1, A-s3, A-s5). Middle row: A-d set (from left to right: A-d1, A-d3, A-d5). Bottom row: B-d set (from left to right: B-d1, B-d3, B-d5). Actual aspect ratios are shown, and all dimensions in the figures are normalized with half channel height $h$. All samples shown in this figure have identical element height distribution, i.e. identical $k_m$ and $\sigma_k$.}
	\end{center}
\end{figure}

 In view of the above definitions, it is now possible to introduce the roughness samples simulated in the present paper. In order to study the effect of roughness morphology, three sets of samples are generated each containing five single samples (total of 14 samples as one geometry is repeated). All 14 samples have identical element height distribution, which allows studying the effect of morphology \textit{independent of roughness height}. Square patches from a number of these samples are shown in figure \ref{fig:samples}. Each row in this figure contains three samples from one set. For the first set, samples with different slopes are generated by `stretching' a single reference sample horizontally. By doing so, the  $k_m/D_m$ ratio varies but $D_m/S_m$ remains constant. The sample with the highest slope (the one shown on the left hand side in the figure) is composed of type A elements introduced above. Obviously, those samples with lower slopes, are composed of laterally stretched type A elements. By comparing samples of the first set, the effect of roughness slope isolated from density can be studied.

The second set (shown in the second row of figure \ref{fig:samples}), on the other hand, can be used to study the effect of density isolated from slope. To generate samples of this set, different numbers of elements with shape A are distributed on the surface, thus here the ratio $k_m/D_m$ remains constant while $D_m/S_m$ varies. The third set is similar to the second, but elements with shape B are used. Each sample from the second set has the same number of elements with one sample in the third, hence same total frontal area (compare samples in the same column in figure \ref{fig:samples}). This allows additionally studying the effect of `element shape' isolated from total frontal area.

The first, second and third sets of samples are labelled A-s, A-d, and B-d (the first letter indicates the element type and the second the property that is varied in this set: s for slope, d for density). The five samples in each set are additionally numbered, decreasing in slope/density, from 1 to 5. It should be mentioned that the parameters are so tuned that all samples with the same number have the same total frontal areas.

Table \ref{table:samples1} lists all the artificial samples described above. As mentioned before, $k_m$ and $\sigma_k$ of all samples are the same when normalized with channel half height $h$. With the knowledge of these two quantities, hence the probability density function (PDF) of element heights, one can find an element height $k$ from which 95\% of all elements are shorter. This size is denoted by $k_{95}$ and can be considered as the `statistical' size of the largest roughness elements within one sample. This is an important quantity as it is suggested in the literature that the largest roughness peak height is possibly the most relevant length scale in determining the equivalent roughness. Other statistical properties listed in the table are $R_q$, $Sk$ and $Ku$, which are root-mean-square, skewness and kurtosis of the PDF of surface elevation , respectively. Additionally, $\lambda_f$ stands for the `frontal solidity' defined as total frontal projected area of roughness per unit bottom plane area. In the last column of figure \ref{fig:samples}, the values of equivalent sand-grain roughness $k_s$ calculated based on the simulation results are displayed. Details of computation of $k_s$ are explained in section 4.

\begin{table}[h]
	\caption{\label{table:samples1} Properties of the artificial surface samples. In all samples $k_m=0.13h$, $k_{95}=0.19h$ and $\sigma_k=0.28k_m$. Samples A-s1 and A-d1 are identical. 
	}	
	\begin{tabular}{p{0.1\textwidth} p{0.1\textwidth} p{0.1\textwidth} p{0.1\textwidth} p{0.07\textwidth} p{0.07\textwidth} p{0.07\textwidth} p{0.07\textwidth} p{0.07\textwidth}} 
		\hline\hline
		Sample  & $k_m/D_m$ & $D_m/S_m$ & $R_q/k_m$ & $Sk$ & $Ku$ & $\lambda_f$ & $k_s/k_{95}$\\ 
		\hline
		$\textrm{A-s1}$  & $1.10$ & $0.54$ & $0.36$ & $0.76$  & $2.8$ & $0.39$ & $2.36$ \\
		$\textrm{A-s2}$  & $0.85$ & $0.54$ & $0.36$ & $0.78$  & $2.9$ & $0.29$ & $2.26$ \\
		$\textrm{A-s3}$  & $0.58$ & $0.54$ & $0.37$ & $0.80$  & $3.0$ & $0.20$ & $1.88$ \\
		$\textrm{A-s4}$  & $0.42$ & $0.54$ & $0.37$ & $0.78$  & $2.9$ & $0.14$ & $1.62$ \\
		$\textrm{A-s5}$  & $0.29$ & $0.54$ & $0.37$ & $0.80$  & $2.9$ & $0.10$ & $1.04$ \\
		\hline
		$\textrm{(A-d1}$  & $1.10$ & $0.54$ & $0.36$ & $0.76$  & $2.8$ & $0.39$ & $2.36)$ \\
		$\textrm{A-d2}$  & $1.10$ & $0.44$ & $0.35$ & $1.23$  & $3.7$ & $0.29$ & $2.63$ \\
		$\textrm{A-d3}$  & $1.10$ & $0.36$ & $0.31$ & $1.81$  & $5.6$ & $0.19$ & $2.73$ \\
		$\textrm{A-d4}$  & $1.10$ & $0.31$ & $0.27$ & $2.25$  & $7.5$ & $0.14$ & $2.58$ \\
		$\textrm{A-d5}$  & $1.10$ & $0.26$ & $0.24$ & $3.01$  & $12.1$ & $0.10$ & $2.14$ \\
		\hline
		$\textrm{B-d1}$  & $1.10$ & $0.51$ & $0.47$ & $1.15$  & $2.8$ & $0.39$ & $2.53$ \\
		$\textrm{B-d2}$  & $1.10$ & $0.44$ & $0.43$ & $1.59$  & $4.1$ & $0.29$ & $2.86$ \\
		$\textrm{B-d3}$  & $1.10$ & $0.36$ & $0.38$ & $2.31$  & $7.1$ & $0.19$ & $2.96$ \\
		$\textrm{B-d4}$  & $1.10$ & $0.31$ & $0.32$ & $2.77$  & $9.5$ & $0.14$ & $2.70$ \\
		$\textrm{B-d5}$  & $1.10$ & $0.26$ & $0.27$ & $3.56$  & $14.7$ & $0.10$ & $2.21$ \\
		\hline\hline\\
	\end{tabular}
	\\
\end{table}

As mentioned before, the effect of roughness height independent of roughness morphology is also to be studied in the present paper. To generate geometries with the same morphology but different characteristic heights, it is enough to scale any of the samples discussed above proportionally in all directions . This is done for a number of selected samples as will be discussed in detail in section 4.2. These \textit{scaled} samples are geometrically similar to those shown in table \ref{table:samples1}, hence are not added to the table.
\subsection{\label{sec:samples2}Realistic roughness}
\begin{figure}[h!]
	\begin{center}
		\includegraphics*[trim={0.5cm 2.0cm 0.0cm 2.4cm},clip,width=0.5\textwidth]{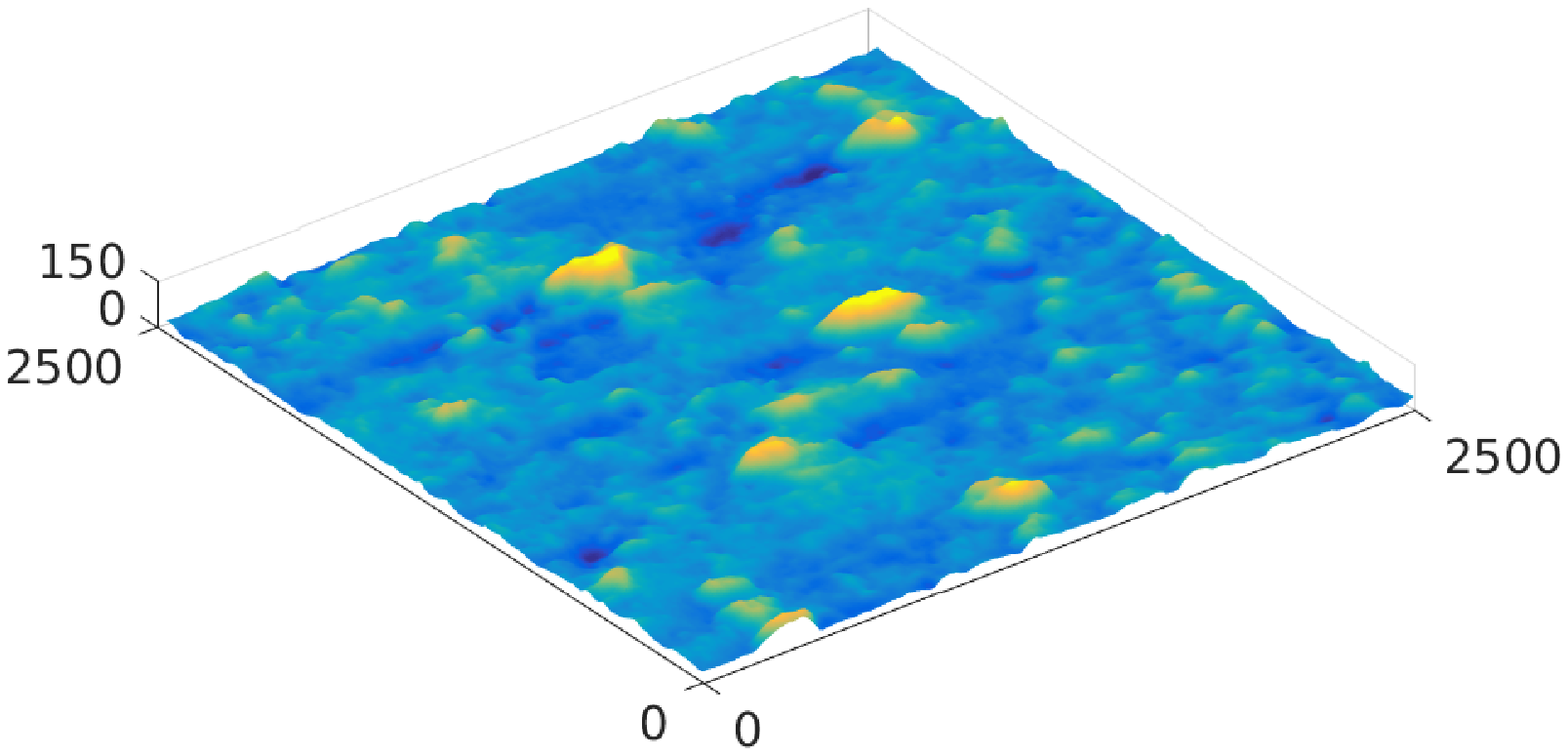}
		\includegraphics*[trim={0cm 0.0cm 0cm 1.4cm},clip,width=0.38\textwidth]{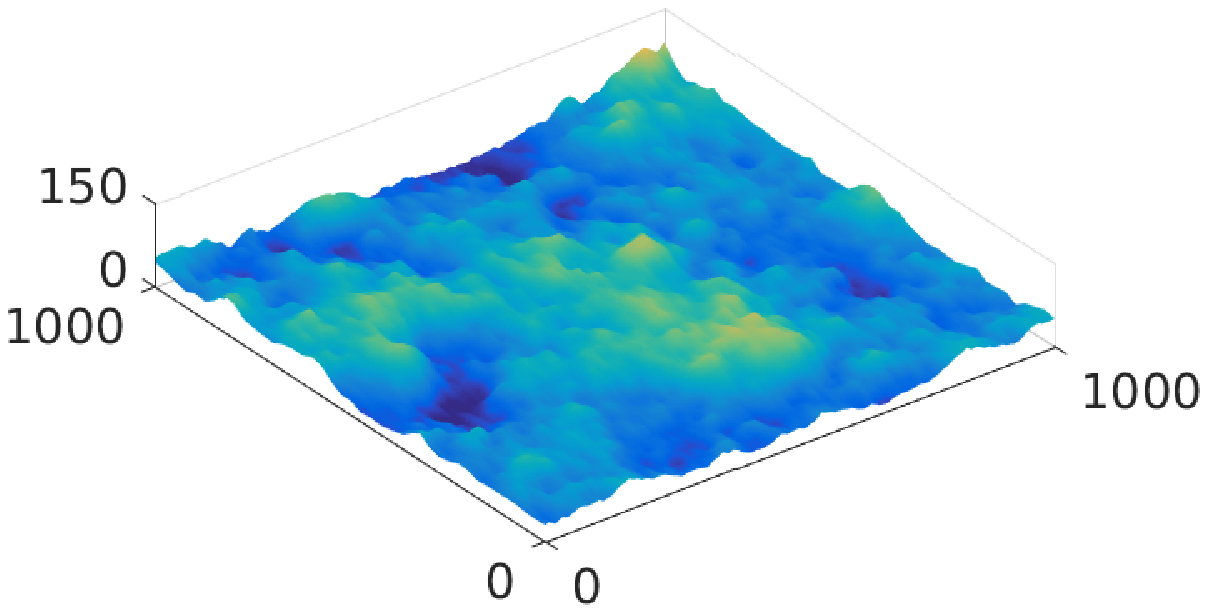}
		\includegraphics*[trim={1.0cm 0.2cm 1.0cm 0cm},clip,width=0.75\textwidth]{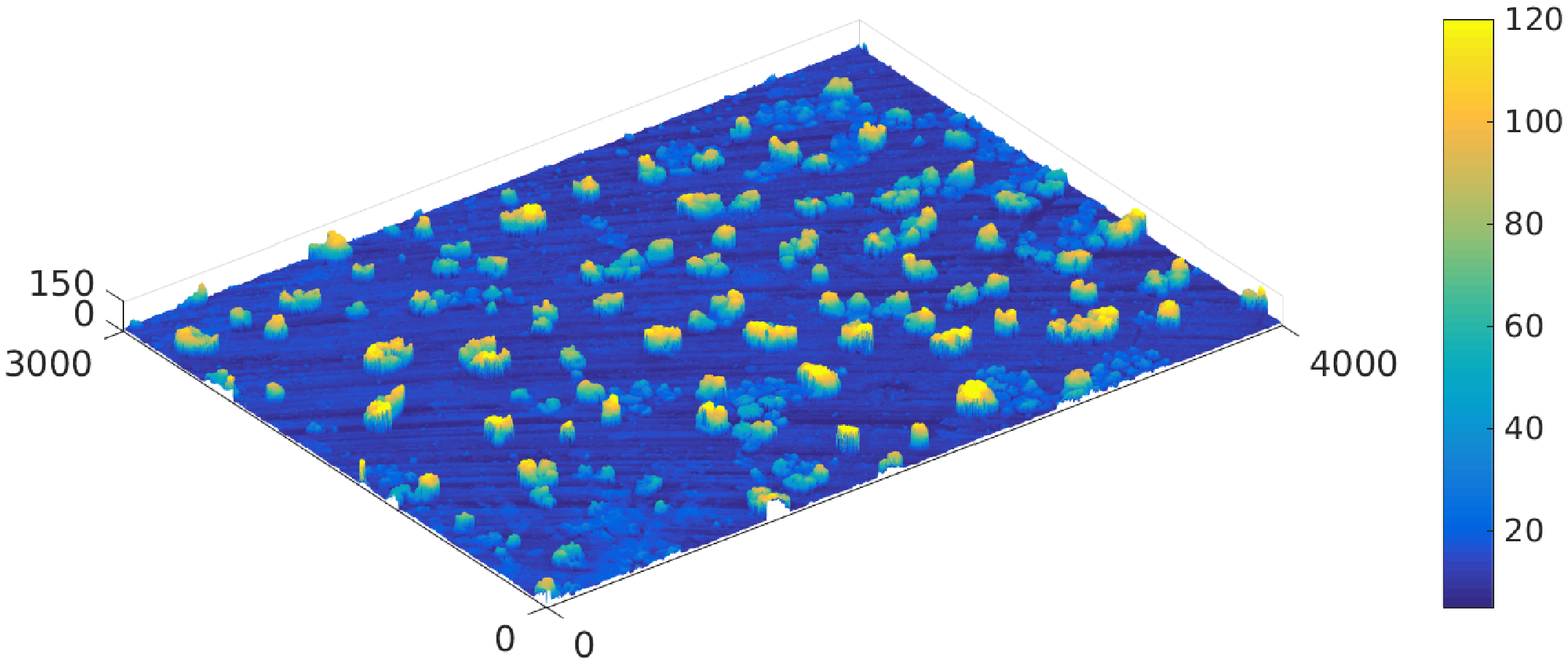}
		\caption{\label{fig:samples2} Realistic surface samples. Top right: GT1. Top left: GT2. Bottom: IC (reproduced from \cite{Forooghi18}). All dimensions are in $\mu m$. Colour bar indicates surface elevation from the lowest valley. Same color bar applies to all maps.}
	\end{center}
\end{figure}

Realistic roughness geometries are used in the present paper to complement the results obtained from the artificial ones. Two roughness geometries, based on surface maps of real turbine blades, are simulated for the first time in this paper. These samples, labelled GT1 and GT2, are displayed in figure \ref{fig:samples2}.  Forooghi et al. \cite{Forooghi18} used DNS to study two roughness geometries based on surface scans of a deposited piston head in a direct injection gasoline IC-engine. Those results are also used in section 4.2 of the present paper to enrich the discussion. One of the surface maps studied in \cite{Forooghi18} is displayed in figure \ref{fig:samples2} (sample s1 in \cite{Forooghi18}). The surface samples from IC-engine are labelled IC in the present paper.

Sample GT1 is extracted from the pressure side of a first-row rotor blade in a civil aircraft engine. Sample GT2 is measured on a surface coated by an air plasma sprayed thermal barrier. The difference in the nature of roughness-formation in the two samples is visible in the surface maps as well as in the statistical properties shown in table \ref{table:samples2}. In sample GT1, and similarly IC, deposition is the dominant mechanism leading to the appearance of more prominent peaks rather than pits. While in GT1 some pits are also present, the roughness in IC is purely formed by particle-like features spread on the bottom plane; what is obvious from the different dominant colours in the surface visualisations. A peak-dominated roughness can be identified from its relatively high value of skewness, which is the case in both samples GT1 and IC as seen in table \ref{table:samples2}. It should be noted that the artificial samples introduced before are also mostly peak-dominated. The roughness in sample GT2 is on the other hand a result of ceramic coating, which is neither peak nor pit-dominated. As a result its skewness is much closer to zero.

All realistic samples (particularly those from gas turbines) have relatively low values of frontal solidity. Sample GT1 has the lowest value of $\lambda_f$ among all simulated samples in this study. For a realistic roughness, it is not straightforward to separate the roughness density from the surface slope as done before for the artificial roughness, and it is also not in the scope of the present paper. Having said that, by a visual comparison of sample GT1 with those in figure \ref{fig:samples} it can be roughly said that this surface features a combination of both sparsity and low slope.

For a realistic roughness, it is also not possible to define a statistical maximum peak height such as $k_{95}$ for artificial roughness. Therefore, we define a peak-to-valley roughness size as the mean of maximum peak-to-valley heights over various subsets of the sample, each subset with $1 \times 1$ $mm^2$ area. This peak-to-valley height is denoted by $R_z$ following the gas turbine community notation.

\begin{table}[h]
	\caption{\label{table:samples2} Properties of the realistic surface samples. }	
	\begin{tabular}{p{0.1\textwidth} p{0.14\textwidth} p{0.1\textwidth} p{0.07\textwidth} p{0.07\textwidth} p{0.1\textwidth} p{0.07\textwidth} p{0.07\textwidth}} 
		\hline\hline
		Sample  & size $[mm^2]$ & $R_q [\mu m]$ & $Sk$ & $Ku$ & $R_z [\mu m]$ & $\lambda_f$ & $k_s/R_z$\\ 
		\hline
		$GT1$  & $2.5 \times 2.5$ & $14.1$ & $1.70$ & $8.4$   & $105$ & $0.09$ & $0.62$ \\
		$GT2$  & $1 \times 1$     & $14.7$ & $0.26$ & $2.7$   & $95$  & $0.17$ & $0.70$ \\
		$IC$   & $3 \times 4$     & $20.9$ & $2.99$ & $11.8$  & $125$ & $0.22$ & $2.50$ \\
		\hline\hline\\
	\end{tabular}
\end{table}

\section{\label{sec:solution}Solution}

\begin{figure}[h!]
	\begin{center}
		\includegraphics*[trim={0cm 0cm 0.0cm 0cm},clip,width=0.5\textwidth]{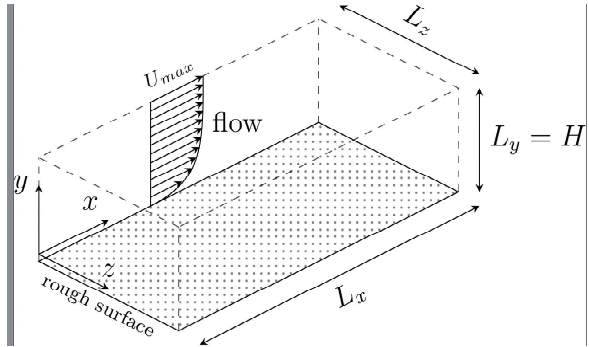}
		\caption{\label{fig:box} Schematic representation of the computational domain. The distance $h$ between the mean height of the rough surface and the top boundary is smaller than the height of computational domain $H$.}
	\end{center}
\end{figure}

A Pseudo-spectral solver SIMSON \cite{Chevalier07} is used for solving the incompressible Navier Stokes and energy equations in a cubic domain as schematically depicted in figure \ref{fig:box}. The streamwise, wall-normal and spanwise coordinates are denoted by $x$, $y$ and $z$. We simulate a half channel (commonly referred to as open-channel DNS in literature \cite{ChanBraun11}) with the wall at the bottom and a symmetry plane at the top. The size of the computational domain in $y$-direction is $H$. The distance between the symmetry plane and the mean roughness elevation, denoted by $h$, is referred to as `effective channel half height' throughout the paper. The solver employs Fourier (in $x$ and $z$-directions) and Chebyshev (in $y$-direction) series for spatial discretization. The Immersed Boundary Method (IBM) based on \cite{Goldstein93} is used to reproduce the exact roughness geometry in the solution. The solver and IBM are thoroughly validated and used in previous publications  \cite{Forooghi18,Forooghi18PRF}, thus in this section only the most important aspects of the solution are briefly explained.

Every simulation is run at a mean streamwise pressure gradient constant in time, which is added to the Navier Stokes equation in form of a constant source term. As a result, wall-shear stress $\tau_w$ can be directly prescribed. Further details on the calculation of wall shear stress is given later. Periodic boundary conditions are applied to the velocity components in streamwise and spanwise directions. No-slip and zero gradient boundary conditions are used for the other two boundaries, which realize a so-called fully developed open-channel numerical set up. In case of an artificial roughness, the roughness elements are attached to the lowest $y$-plane in the computational domain (bottom plane). For the realistic roughness, the lowest valley of the rough surfaces is level with this plane. The rough interface divides the computational domain into the fluid and the solid sub-domains. In the former, Navier Stokes equations are solved normally while in the latter the velocity is forced to vanish by means of IBM creating a no-slip surface within the computational domain.

For the realistic roughness, the surface maps shown in figure \ref{fig:samples2} are used as tiles to cover the bottom plane. The orientation of each tile is randomly determined to keep the stochastic nature of the roughness\footnote{The samples are randomly `mirrored' rather than being rotated as the latter changes the directional properties of the surface in case of an anisotropic roughness.}. Gradual transition is applied at the edges of the tiles to create a smooth, seamless and periodic surface geometry. The number of tiles is determined based on the desired roughness to channel height ratio $R_z/H$. Once this ratio is prescribed, it is possible to calculate the dimensions of the tile in $H$ units. Finally, one can calculate how many tiles are required to cover an approximately $8H \times 4H$ (strwamwise $\times$ spanwise) area. These dimensions guarantee that the computational domain is large enough so that the first and second order one-point statistics are domain-size independent \cite{LozanoDuran14}. Based on this criterion, for GT1, $2 \times 1$ tiles and for  GT2 (similarly for one of the GT1 simulations with a smaller $R_z/H$ ratio), $4 \times 2$ tiles are required.

The energy equation for an incompressible and non-reacting flow with constant thermophysical properties reduces to a passive scalar equation. For thermal boundary conditions, we follow the approach introduced by Kasagi et al. \cite{Kasagi92}. This approach, also known as mixed-type boundary condition, is well posed and most widely used in DNS studies \cite{Piller05}. With the so-called mixed-type boundary condition, the time-averaged wall heat flux $q_w$ is uniform in space and can be found from a prescribed streamwise temperature gradient. In this approach, energy equation is solved with the temperature difference $\theta=T_w-T$ as the dependent variable, where $T$ is the fluid temperature and $T_w$ is the wall temperature at the corresponding streamwise location. Change of variable from $T$ to $\theta$ entails the appearance of a source term in the energy equation, which is equal to the product of streamwise velocity and a prescribed streamwise temperature gradient (see \cite{Piller05}). Temperature difference $\theta$ is by definition zero on the wall. This condition is forced at the solid sub-domain by the IBM similar to the no-slip condition for the velocity. Periodic boundary conditions are used for $\theta$ in streamwise and spanwise directions, which translates to a thermally fully-developed flow. Prantdl number in all simulations is equal to 0.71.

Wall shear stress $\tau_w$ can be calculated based on the global momentum balance from the prescribed mean streamwise pressure gradient $P_x$ \cite{ChanBraun11}: 
\begin{equation}
\tau_w=P_xh
\label{eq:tau_w}.
\end{equation}
The wall shear stress calculated in this way is equal to the total wall resistance force, i.e. the sum of the forces due to the viscous shear stress and pressure drag on the roughness, per unit wall-projected area. Based on this quantity, we define friction velocity $u_\tau=(\tau_w/\rho)^{1/2}$ and viscous length scale $\delta_v=\nu/u_\tau$ to be used for non-dimensionalization in viscous or inner units indicated by a plus superscript $()^+$. Friction Reynolds number is defined as $Re_\tau=u_\tau h/\nu$ or similarly $Re_\tau=h^+$. Friction coefficient and Stanton number can be calculated based on the prescribed wall shear stress $\tau_w$ and mean wall heat flux $q_w$ and the computed bulk quantities. Here $q_w$ is the total wall heat transfer per unit wall-projected area.
\begin{equation}
C_f=\frac{\tau_w}{\rho u_b^2/2} \quad , \quad St=\frac{q_w}{\rho c_p u_b \theta_b}
\label{eq:tau_w}
\end{equation}
Bulk velocity $u_b$ and bulk temperature difference $\theta_b$ can be found from the solved velocity and temperature fields:
\begin{equation}
u_b= \frac{1}{hL_z}\int_{A_c} \overline{u}dA \quad , \quad \theta_b= \frac{1}{u_bhL_z}\int_{A_c} \overline{u}\overline{\theta}dA
\label{eq:tau_w}
\end{equation}
where $A_c$ denotes the free cross section area and $\overline{u}$ and $\overline{\theta}$ are mean streamwise velocity and mean temperature difference, respectively.

The simulations are carried out at three friction Reynolds numbers approximately equal to 500, 380 and 230. In all simulations a grid with $1152\times301\times576$ spectral points (streamwise $\times $ wall-normal $\times$ spanwise) is used. This corresponds to a mesh sizes of $\Delta x^+ = \Delta z^+ = 3.6$ and $\Delta y^+=0.02$ -- $2.7$ in case of the highest friction Reynolds number. A summary of the simulated cases and the corresponding domain sizes are presented in table \ref{table:cases}. A sum of 20 simulations are run for the artificial roughness. This includes one simulation for each of the 14 samples shown in table \ref{table:samples1} at $h^+=500$ and $k_{95}/h=0.19$, thus $k_{95}^+=95$. Additionally for each of the samples As-3 and Ad-3, three simulations are run to study the effect of roughness size, including two simulations at $k_{95}/h=0.19$ but other values of $k_{95}^+$ and one at a different $k_{95}/h$. For each of the artificial samples one simulation is run at $h^+=500$ and $R_z/h=0.19$. For sample GT1, three simulations are added at other values of $R_z/h$ or $R_z^+$ in the same way as described for A-s3 and A-d3.

\begin{table}[h]
	\centering
	\caption{\label{table:cases} Summary of all simulated cases. 
	}	
	\begin{tabular}{p{0.25\textwidth} p{0.15\textwidth} p{0.25\textwidth} } 
		\hline\hline
		Type  & $L_x \times L_z$ & no. of simulations\\ 
		\hline
		Artificial roughness  & $8.0H \times 4.0H$ & $20$ \\
		GT1  & $8.2H \times 4.1H$ & $4$ \\
		GT2  & $7.6H \times 3.8H$ & $1$ \\
		\hline\hline\\
	\end{tabular}
	\\
\end{table}

\section{\label{sec:results}Results and discussion}

\subsection{\label{sec:results1}Effect of surface morphology}
In what follows, we first discuss the effect of surface slope by presenting the results of A-s samples at $h^+=500$ and $k_{95}/h=0.19$ (section 4.1.1), and then the effect of roughness element density and shape by presenting the results of A-d and B-d samples also at $h^+=500$ and $k_{95}/h=0.19$ (section 4.1.2). Finally, we compare the Reynolds analogy factor ratio $RA/RA_0$ of all the above mentioned samples as well as the two realistic samples at $h^+=500$ and $R_z/h=0.19$, and  evaluate two commonly-used roughness metrics --  so called density parameters -- for their capability in correlating this ratio (section 4.1.3). Since the friction Reynolds number is kept constant in this section, the bulk velocity is not constant due to different friction coefficients for different surfaces. The bulk Reynolds number (based on $u_b$ and $4h$) varies within the range of 14,000 to 23,000 for all the cases discussed in this section. For the calculation of the smooth-wall Reynolds analogy factor $RA_0$, throughout the paper, Stanton number and friction coefficient are calculated based on the Dittus-Boelter and Dean's correlations at the same bulk Reynolds number as the rough surface of interest. For application of Dittus-Boelter correlation, $4h$ is used as the hydraulic diameter. 
\subsubsection{\label{sec:results11}Effect of surface slope}
\begin{figure}[h!]
	\begin{center}
		\includegraphics*[trim={1cm 0.5cm 0.5cm 1cm},clip,width=0.49\textwidth]{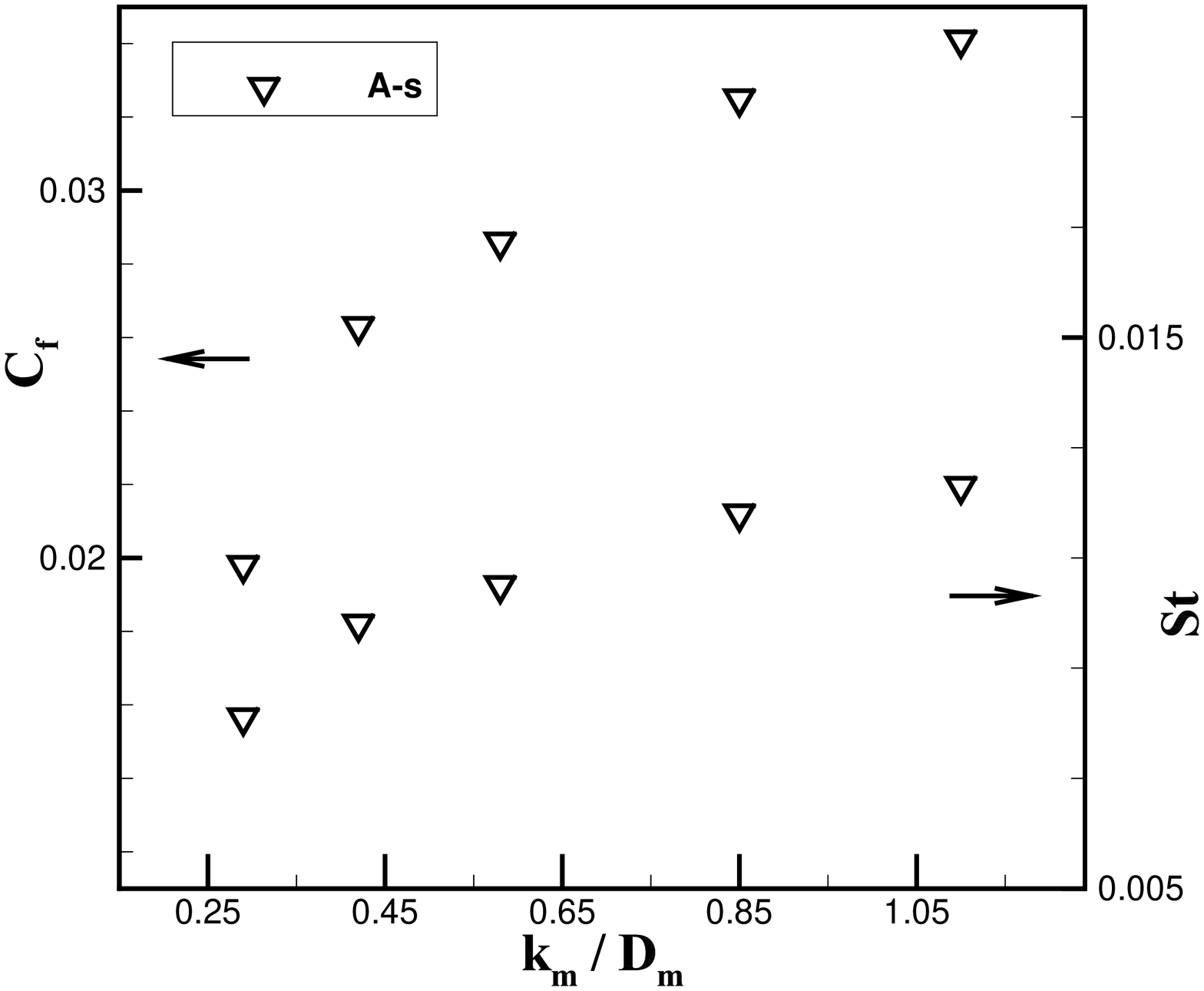}
		\includegraphics*[trim={1cm 0.5cm 0.5cm 1cm},clip,width=0.49\textwidth]{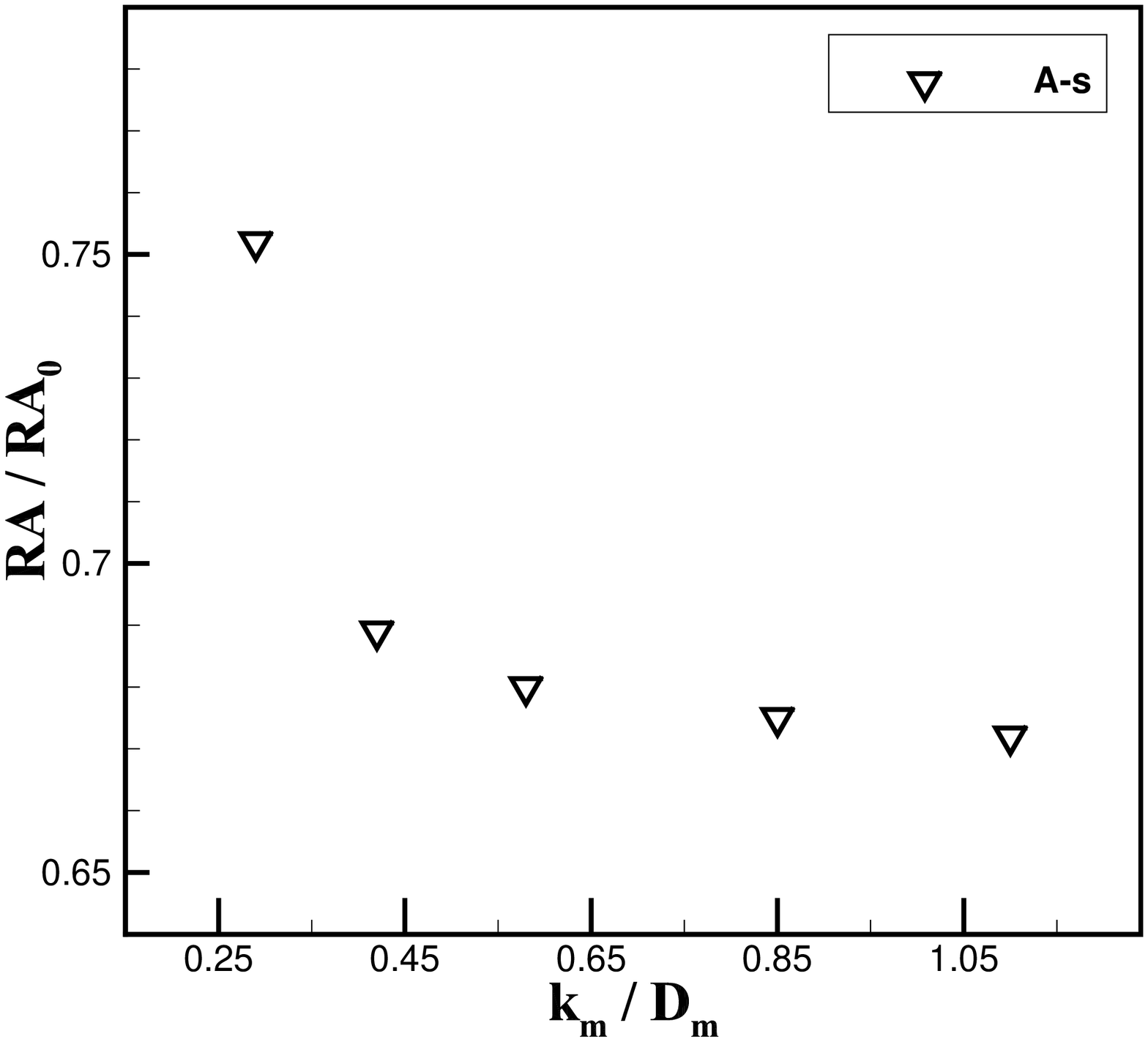}
		\caption{\label{fig:slope} Variation of $C_f$ and $St$ (left) and $RA/RA_0$ (right) against the mean roughness element height to diameter ratio $k_m/D_m$ for the artificial roughness samples in the A-s set. With an increase in $k_m/D_m$ surface slope increases. For all data points, $D_m/S_m=1.1$, $k_{95}^+\cong95$ and $k_{95}/h=0.19$.}
	\end{center}
\end{figure}

Figure \ref{fig:slope} shows the variations of $C_f$ and $St$ (left) and $RA/RA_0$ (right) as a function of $k_m/D_m$ or in other words as a function of surface slope. It is observed that with an increase in the surface slope, both friction coefficient and Stanton number increase, which is a physically reasonable outcome since with a decreasing slope, the rough surface tends to a smooth one, and expectedly, the momentum and heat transfer augmentation due to roughness vanish. The augmentation of $C_f$ with the surface slope is however stronger than that of $St$, which means that the ratio $RA/RA_0$ decreases when surface slope increases. For all rough surfaces studied in this paper, $RA/RA_0$ is smaller than 1. As discussed in the section 1, it is a known phenomenon stemming from the fact that the pressure (form) drag on the rough surface plays a dominant role in the augmentation of friction coefficient, a  mechanism that is absent in heat transfer. The decrease in $RA/RA_0$ with slope for the points shown in figure \ref{fig:slope} is rather moderate -- except for the two data points with the smallest values of $k_m/D_m$ -- and saturates at the high values of $k_m/D_m$. It can be concluded that $RA/RA_0$ shows little sensitivity to the slope for the steeper surfaces, but sharply tends to 1 below a certain threshold. Note that $RA/RA_0$ is by definition 1 for $k_m/D_m=0$ (i.e., a smooth wall). 
\subsubsection{\label{sec:results12}Effect of roughness element density and shape}
\begin{figure}[h!]
	\begin{center}
		\includegraphics*[trim={1cm 0.5cm 0.5cm 1cm},clip,width=0.49\textwidth]{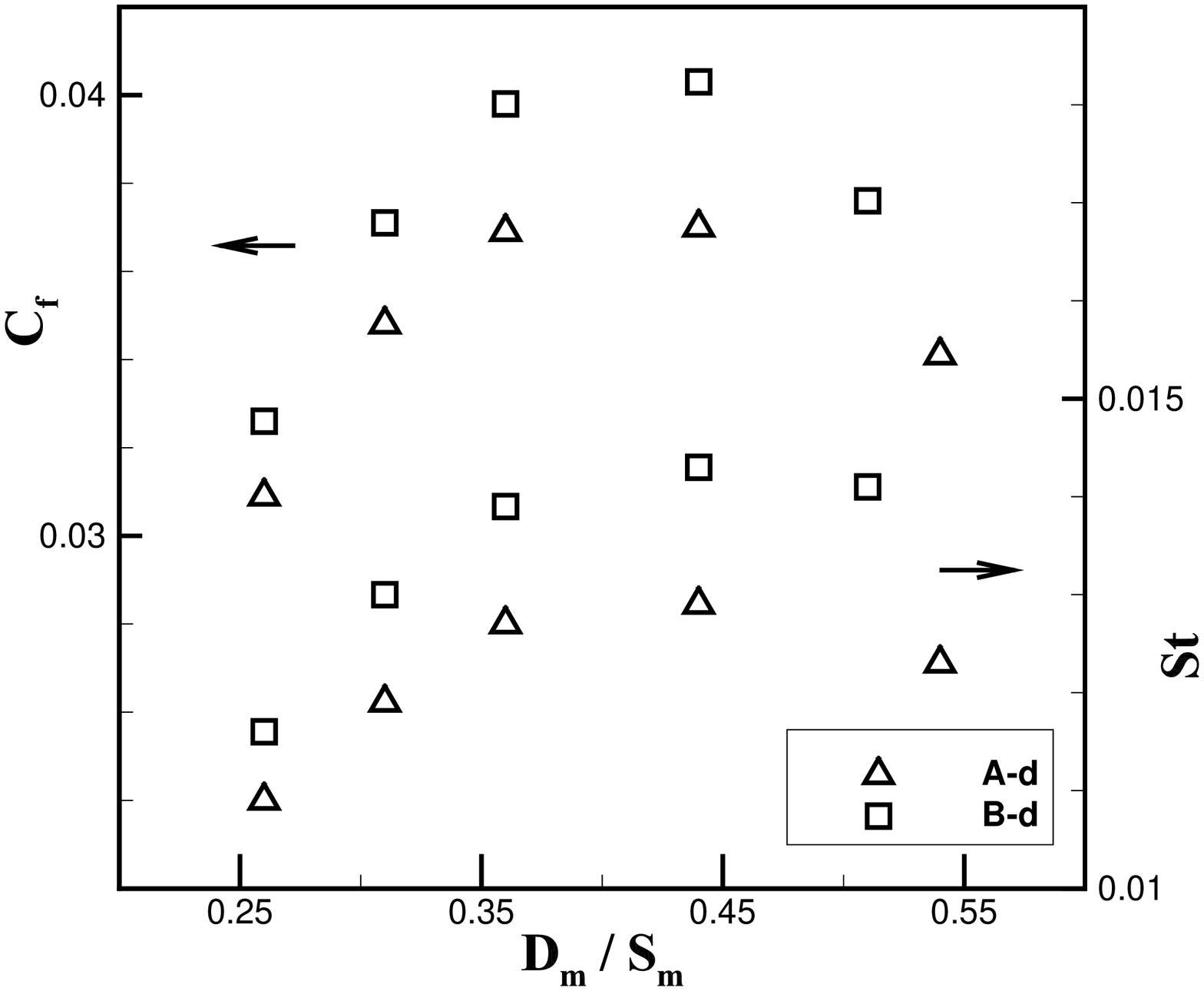}
		\includegraphics*[trim={1cm 0.5cm 0.5cm 1cm},clip,width=0.49\textwidth]{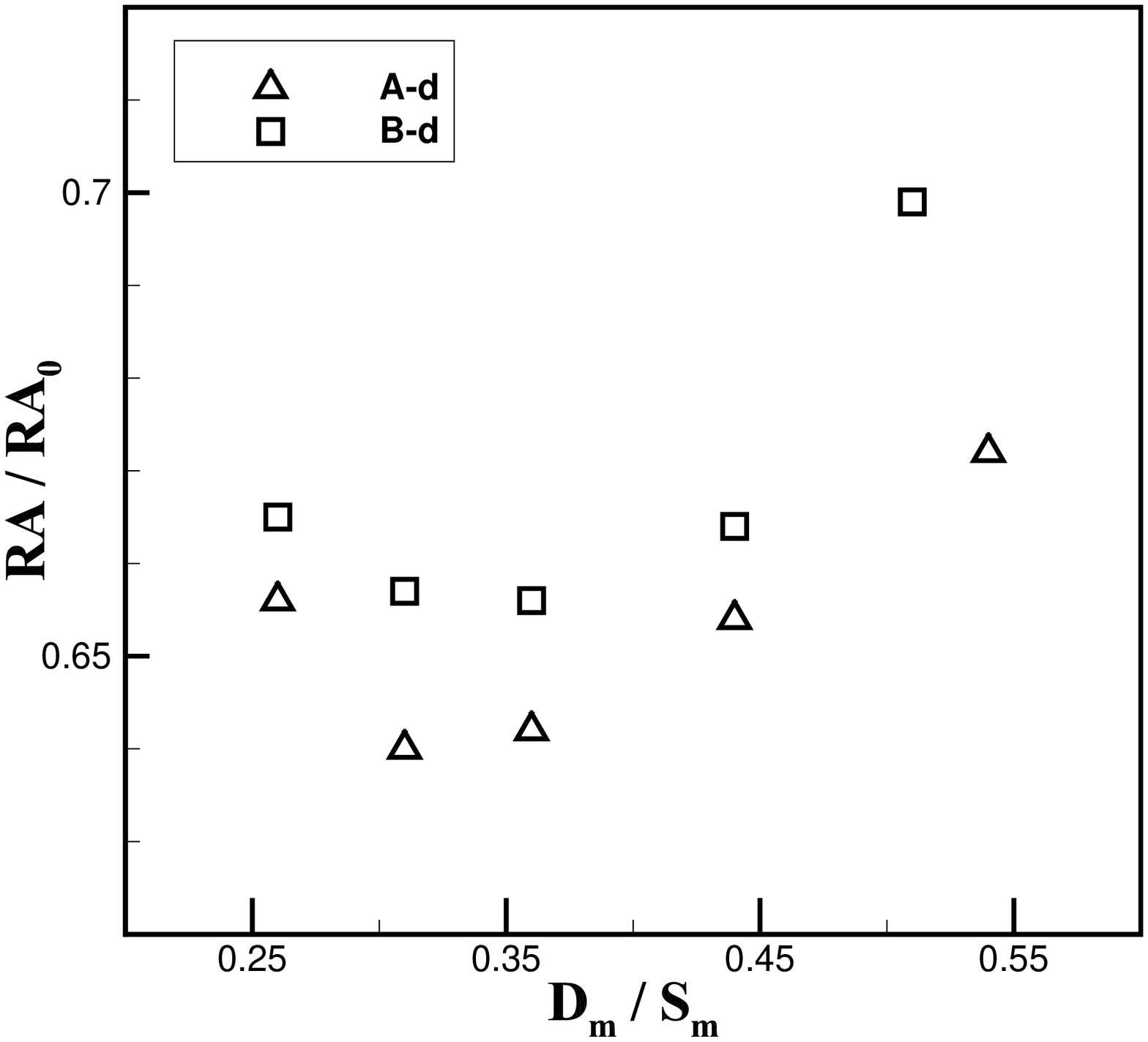}
		\caption{\label{fig:density} Variation of $C_f$ and $St$ (left) and $RA/RA_0$ (right) against the mean roughness element diameter to spacing ratio $D_m/S_m$ for the artificial roughness samples in the A-d and B-d sets. With an increase in $D_m/S_m$ roughness density increases. For all data points, $k_m/D_m=0.54$, $k_{95}^+\cong95$ and $k_{95}/h=0.19$. In both plots from left to right surface slope increases.}
	\end{center}
\end{figure}

Figure \ref{fig:density} shows the variations of $C_f$ and $St$ (left) and $RA/RA_0$ (right) as a function of $D_m/S_m$ or in other words as a function of roughness density. Two sets of artificial samples, A-d and B-d, are represented in this figure. As mentioned before, the geometries are so designed that the two samples from two sets with roughly the same $D_m/S_m$ also have the same total frontal area. Therefore, the effect of the shape of roughness elements  independent of their density and frontal area can be additionally studied in this figure. It is observed that both $C_f$ and $St$ values as well as the ratio $RA/RA_0$ are larger for the B-d sample compared to A-d, indicating that the shape of roughness elements has an influence on flow and heat transfer irrespective of their frontal area and spacing. Comparing the roughness elements of type A and B, the latter are wider on top, i.e., where the fluid has higher velocities. As a result, the elements of B type block the flow more effectively despite having the same frontal area as those of A type. Higher Reynolds analogy factor of the B type roughness elements, on the other hand, is possibly due to their flatness on top, that can enhance the contribution of viscous transport.

Regarding the trends of $C_f$ and $St$ with roughness density, for either roughness element shape, both quantities show a peak when $D_m/S_m$ varies. This peaking behaviour is already known, at least for the momentum transfer, to be a result of a mechanism often referred to as `sheltering' of roughness elements \cite{MillwardHopkins11}. When the roughness becomes dense, some roughness elements lie fully or partly in the low velocity wake regions behind the upstream elements (sheltered), which leads to a reduction of friction coefficient despite an increase in total surface area. It is reasonable to assume that a mechanism similar to sheltering leads to a decrease in heat transfer coefficient beyond a certain density as the roughness elements become surrounded by the low speed, low temperature fluid in the wake of the upstream element rather than being in contact with high velocity, high temperature free flow. We refer to this mechanism as `thermal sheltering'. It should be noted that due to the lack of the pressure term in the energy equation, momentum and thermal sheltering do not necessarily behave identically. It is observed in \ref{fig:density} (left) that the peak due to thermal sheltering occurs at slightly higher densities, although determining the exact locations of the peaks would require more data points.

It is observed in figure \ref{fig:density} (right) that the ratio $RA/RA_0$ has a minimum, which is a sign that the friction coefficient grows (falls) faster with roughness density before (after) the sheltering point. Considering that both extremely high and low roughness densities ($D_m/S_m \rightarrow 0$ or $\infty$) correspond to a smooth wall, where $RA/RA_0=1$, having a minimum for the Reynolds analogy factor is justified. As mentioned before, momentum sheltering occurs earlier than thermal sheltering. As a result, the minimum in $RA/RA_0$ is slightly shifted towards lower values of density as observed in the figure.

\subsubsection{\label{sec:results13}An assessment of two roughness parameters}
\begin{figure}[h!]
	\begin{center}
		\includegraphics*[trim={0cm 0cm 1.5cm 1cm},clip,width=0.49\textwidth]{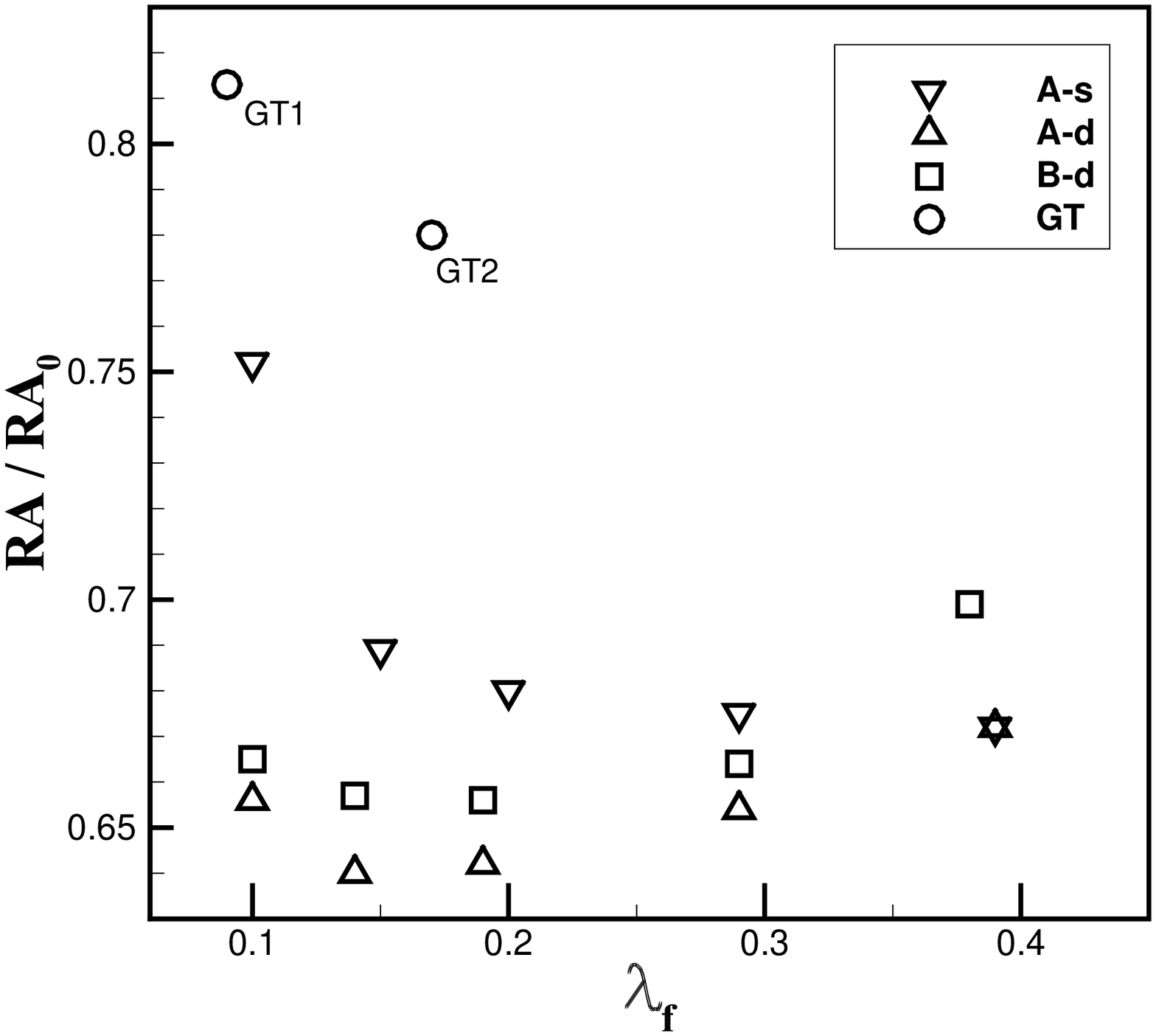}
		\includegraphics*[trim={0cm 0cm 1.5cm 1cm},clip,width=0.49\textwidth]{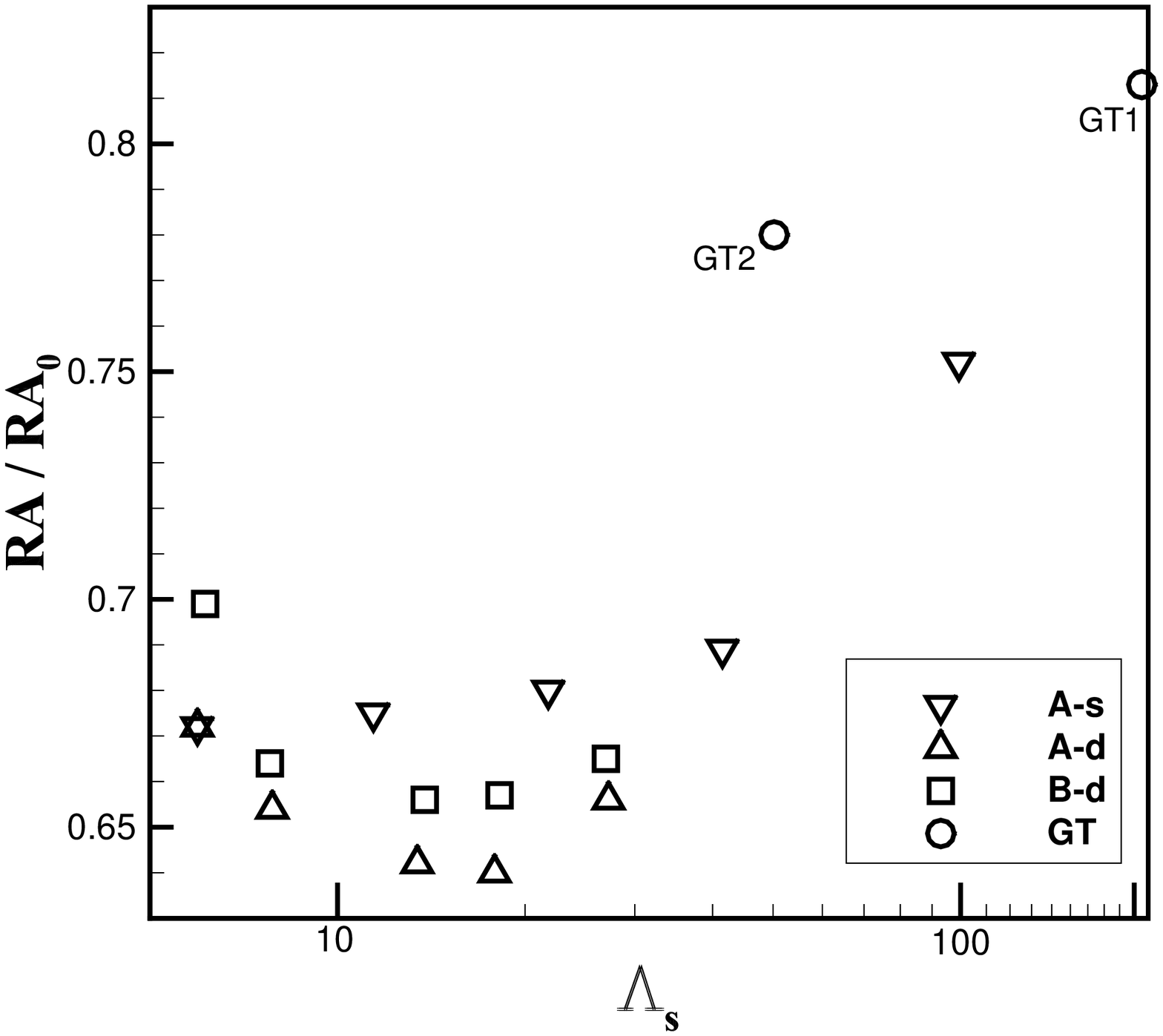}
		\caption{\label{fig:parameters} Variation of $RA/RA_0$ against frontal solidity $\lambda_f$ (left) and the roughness parameter $\Lambda_s$ (right) for all artificial samples with $k_{95}^+\cong95$ and $k_{95}/h=0.19$ as well as the realistic gas turbine roughness samples with $R_z^+\cong95$ and $R_z/h=0.19$. The two overlapping data points from the A-s and A-d sets are identical (A-s1 and A-d1).}
	\end{center}
\end{figure}

So far we used geometrical proportions of the roughness elements to independently study the effect of different morphological properties (slope, density and shape) on heat transfer. Obviously, these proportions can only be defined for a roughness with simple and isolated roughness elements, and thus need to be replaced by `generalized' roughness metrics to be used in the correlations aimed at industrial roughness. Although roughness metrics are not in the focus of the present paper, we find it instructive to assess the performance of two widely used parameters in correlating the present DNS results. For this purpose Reynolds analogy factor ratios from all simulations discussed in sections 4.1.1 and 4.1.2 as well as those of the realistic samples GT1 and GT2 are plotted against the two roughness metrics of interest in figure \ref{fig:parameters}. The physical height of roughness -- $k_{95}$ for the artificial and $R_z$ for the realistic samples -- is the same in both inner and outer units for all cases presented in this figure.

The first roughness parameter discussed here is the frontal solidity $\lambda_f$, the ratio of total frontal-projected surface area to the bottom plane surface area. This parameter, already introduced in section 1, can be easily linked to the pressure drag due to roughness and is widely discussed in the fluid mechanics community \cite{Jimenez04,PlacidiGanapati15}. A fact that needs to be reminded about the frontal solidity is that it monotonically increases with an increase in both density and slope of roughness as they are defined in this paper. In other words, both geometric transformations shown in figure \ref{fig:schematic} lead to a reduction in $\lambda_f$, while they do not necessarily have the same effect on the physics of flow and heat transfer as discussed in the previous sections. This is, in our view, a weakness of this parameter, as will be explained later.

The second roughness parameter originates from the experimental studies of Sigal and Danberg \cite{SigalDanberg90} and van Rij et al. \cite{vanRij02}. The latter authors defined a generalized form of roughness parameter as
\begin{equation}
\Lambda_s=\left(\frac{S}{S_f}\right)\left(\frac{S_f}{S_s}\right)^{-1.6}
\label{eq:vanRij}.
\end{equation}
Where $S$, $S_f$ and $S_s$ stand for bottom surface area, total frontal-projected surface area and total windward, wetted surface area, respectively. Indeed parameter $\Lambda_s$ is inverse of frontal solidity corrected with a so-called roughness shape parameter $\left(S_f/S_s\right)^{-1.6}$.

Figure \ref{fig:parameters} (left) shows the ratio $RA/RA_0$ plotted against $\lambda_f$. Notably, the two realistic samples in this plot (GT1/2) have larger values of $RA/RA_0$ at the same $\lambda_f$ when compared to the artificial samples. A possible explanation is that, unlike artificial roughness, in a realistic roughness very small surface features (and possibly measurement noise) can exist that may be too small to meaningfully interact with the flow but nevertheless alter some of the surface statistics. As a mater of fact, frontal solidity is directly related to the surface height derivative ($d\tilde{y}/dx$ where $\tilde{y}$  is the surface height) \cite{Bons02} a variable that is sensitive to the presence of small-scale noise. The outcome of surface statistics being `contaminated' by the small scale roughness features is an increase in the frontal solidity. Consequently the data points from realistic samples may not follow the same trend as those of artificial samples when plotted against frontal solidity as they are shifted towards higher values of $\lambda_f$. For the sample GT2, it can be additionally pointed out that, unlike all other geometries studied in the present work, it is not a peak dominated roughness (skewness nearly zero). Consequently, this sample can possess somewhat different transport properties. Expectedly, a non-skewed roughness such as GT2 has a larger $RA$ than a peak dominated roughness as the presence of the protruding peaks in the latter enhances the contribution of form drag and reduces the Reynolds analogy factor.

Considering the data points of the artificial samples in figure \ref{fig:parameters} (left), it is observed that the A-s samples show a monotonically decrease with $\lambda_f$ while A-d and B-d reach a minimum. These two trends are already linked, in the previous sections, to the variations in roughness element slope or density, but cannot be distinguished by use of frontal solidity since frontal solidity varies monotonically with both slope and density. When $RA/RA_0$ is plotted against $\Lambda_s$ in figure \ref{fig:parameters} (right), however, the quality of correlation is to some extent improved. Similar to $\lambda_f$, the parameter $\Lambda_s$ varies monotonically -- but inversely -- with both slope and density, but due to the correction by the shape parameter, it grows much faster with a decrease in the slope. As a result, the data points with very low slope, those with the highest $RA/RA_0,$ are distinguished from the rest of the data points. It seems that the data point from sample GT1 also falls into the same correlation with the artificial data points when plotted against $\Lambda_s$, which is due to the fact that the shape parameter to some degree smoothens the effect of small roughness features. The GT2 data point, however, still shows a systematic increase compared to the rest of data, which, as explained above, can be due to the different nature of this surface.

It is notable that although the ratio $RA/RA_0$ has meaningful trends with all morphological parameters discussed so far, its sensitivity is relatively low except when the roughness is extremely sparse or the slope is very low, i.e., when $\Lambda_s>50$ in figure \ref{fig:parameters} (right). It can be seen, as a consequence, that although the data points on the left hand side of this limit follow different trends, it does not lead to a high scatter and the quality of the correlation with $\Lambda_s$ is still acceptable.

\subsection{\label{sec:results2}Effect of roughness height}
So far we discussed the results of different roughness samples at fixed physical roughness height in both inner and outer units. It this section we select three roughness geometries to study the effect of a variation in roughness height. Two artificial samples, A-s3 and A-d3, and one realistic sample, GT1, are selected for this purpose. The results of the IC samples from \cite{Forooghi18} are also used to enrich the discussion when appropriate.

To keep a uniform notation during this section, we use symbol $k_z$ to indicate the physical roughness height. For an artificial roughness $k_z\equiv k_{95}$ and for a realistic roughness $k_z\equiv R_z$. It should be noted that both $k_{95}$ and $R_z$ are similarly a statistical measure of maximum peak-to-valley roughness height even though they are defined differently because of the inherent difference between the realistic and isolated roughness. It is worth mentioning that if one estimates $R_z$ for the artificial samples, using the conversion of 1 $mm$ in plus units, the calculated values of $R_z$ and $k_{95}$ for the samples A-s3 and A-d3 are the same within $\pm 5 \%$ error confirming that the two parameters are closely related. 

In section 4.2.1, we keep $k_z/h$ constant at $0.19$ and vary $k_z^+$ by prescribing different values of friction Reynolds number $h^+$. Three values of $k_z^+\cong 95$, $72$ and $44$ ($h^+\cong500$, $380$ and $230$) are examined for each geometry (for IC the range is different). In section 4.2.2, $k_z/h$ is varied at constant $k_z^+\cong44$. Two values of $k_z/h\cong0.19$ and $0.09$ are examined for each geometry ($h^+\cong230$ and $500$).

Before discussing the results, it is useful to have a brief discussion on the hydrodynamic regime of the simulated flows. It is well established that the logarithmic region of mean velocity profile over a rough wall undergoes a downward shift $\Delta U^+$, often referred to as the `roughness function', compared to the standard logarithmic law of the wall. In a fully rough regime this downward shift has a logarithmic relation with the inner scaled roughness height. A result of this logarithmic relation is the friction factor being invariant with the bulk Reynolds number. Nikuradse \ref{eq:Nikuradse} reported this logarithmic relation for the first time in his sand-grain roughness experiments
\begin{equation}
\Delta U^+ = B-8.48+\frac{1}{\kappa}\textrm{ln}(k^+_s)
\label{eq:Nikuradse}
\end{equation}
where $\kappa$ and $B$ are von K{\'a}rm{\'a}m constant and the log-law intercept for a smooth wall. Size of sand-grain in the Nikuradse experiments is denoted by $k_s$. For any arbitrary surface, the physical roughness height $k_z$ can be scaled so that the roughness function collapses into Nikuradse's data; the scaled roughness height is the `equivalent sand-grain roughness height', also denoted by $k_s$ hereinafter. Equivalent sand-grain roughness is in general case not a physical size but the `hydrodynamic' scale of the roughness \footnote{Here we are referring to the so-called k-type roughness; a majority of the roughness faced in real applications belong to this category.}.

\begin{figure}[h!]
	\begin{center}
		\includegraphics*[trim={0cm 0cm 1.5cm 1cm},clip,width=0.6\textwidth]{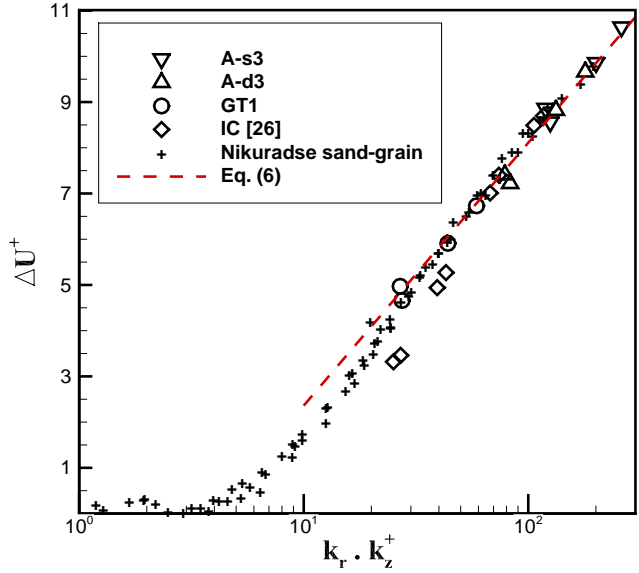}
		\caption{\label{fig:fully} Variation of roughness function with inner scaled equivalent sand-grain roughness ($k_s=k_r\cdot k_z.$). For sand-grain roughness $k_r=1$ by definition.}
	\end{center}
\end{figure}

In view of the above, it is possible to determine if the flow is `fully rough' and determine the value of $k_s$ by plotting the computed values of roughness function against inner scaled roughness height and compare to the Nikuradse data. This plot is displayed for the present data in figure \ref{fig:fully}. The roughness height $k_z$ for each geometry is scaled with a factor $k_r$ to fit into Eq. \ref{eq:Nikuradse}, therefore, $k_s=k_r k_z$. Details on the calculation of roughness function can be found in \cite{Forooghi18}. In the present paper the value of $\kappa=0.38$ is used for the von K{\'a}rm{\'a}m constant in a channel which is the value suggested by Lozano-Duran et al. \cite{LozanoDuran14}.

It is observed in figure \ref{fig:fully} that all the present data collapse satisfactorily with the fully rough asymptote, indicating a fully rough flow over the entire range of roughness sizes. Only the IC samples at the lower values of $k_s^+$ visibly deviate from this line. It should be noted that the onset of the fully rough regime is not determined by a universal threshold and can occur at different values of $k_s^+$ depending on the geometry \cite{FlackSchultz10}, a fact that is also observed here. The values of $k_r$ obtained for the samples in figure \ref{fig:fully}, as well as all other samples, are reported in tables \ref{table:samples1} and \ref{table:samples2}. This is done for the other samples with the assumption that they have reached the fully rough regime at $k_z^+= 95$, which is a reasonable assumption based on figure \ref{fig:fully}.

\subsubsection{\label{sec:results21}Effect of roughness height in inner units}

\begin{figure}[h!]
	\begin{center}
		\includegraphics*[trim={0cm 0cm 1.5cm 1cm},clip,width=0.6\textwidth]{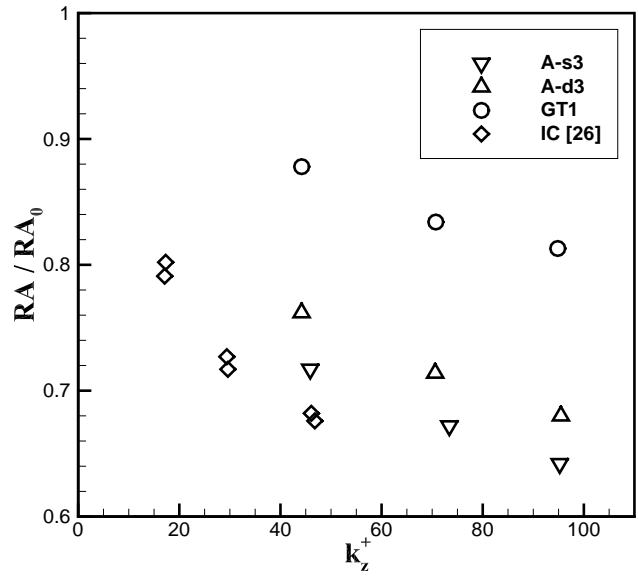}
		\caption{\label{fig:kplus} Variation of $RA/RA_0$ with inner scaled roughness height $k_z^+$ at constant outer scaled roughness height $k_z/h$ for three roughness samples A-s3, A-d3 and GT1. Data from \cite{Forooghi18} are also included in the plot.} 
	\end{center}
\end{figure}

Figure \ref{fig:kplus} shows the variation of Reynolds analogy factor ratio $RA/RA_0$ with the inner scaled physical roughness height $k_z^+$ at constant $k_z/h$. Data points from \citep{Forooghi18}, in which $k_z/h$ is constant are also added to the figure. A clear decreasing trend is observed for all roughness samples. As roughness size increases with respect to the viscous length scale, viscous transport becomes less and less significant, which translates to a decrease in the Reynolds analogy factor. As expected, sample $IC$, the one that has been shown to be partly in the transitionally rough regime, is the one with the fastest decrease in Reynolds analogy factor. Notably, even in the `hydrodynamically' fully rough flows, $RA$ drops with an increase in $k^+_z$. As will be revealed in section \ref{sec:results3}, Reynolds analogy factor most likely saturates when roughness height grows further.

It is revealed in figure \ref{fig:kplus} that the $RA/RA_0$ values from different samples do not collapse when plotted against `physical' roughness height, which is clearly due to the dependence on the morphological parameters or, in other words, the fact that the physical roughness height is not representative enough of the hydrodynamic effect of the roughness. It will be discussed in section \ref{sec:results3} that use of equivalent sand-grain roughness height can lead to a much better collapse of the data points.

\subsubsection{\label{sec:results22}Effect of roughness height in outer units}
\begin{figure}[h!]
	\begin{center}
		\includegraphics*[trim={0cm 0cm 1.5cm 1cm},clip,width=0.6\textwidth]{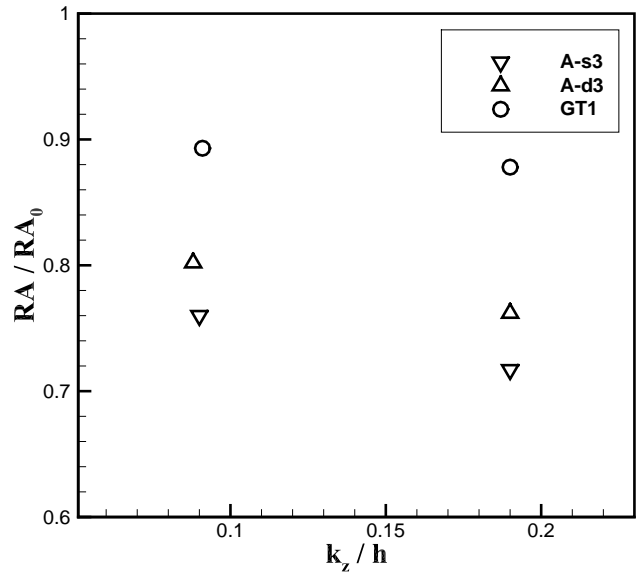}
		\caption{\label{fig:kh} Variation of $RA/RA_0$ with outer scaled roughness height $k_z/h$ at constant outer scaled roughness height $k_z^+\cong44$ for three roughness samples A-s3, A-d3 and GT1.}
	\end{center}
\end{figure}

Figure \ref{fig:kh} shows the variation of Reynolds analogy factor ratio $RA/RA_0$ with the outer scaled physical roughness height $k_z/h$ at constant $k_z^+$. It is observed that for the cases under investigation, $RA/RA_0$ shows a decreasing trend with $k_z/h$ independent of the inner scaling of the roughness size. It should be recalled that the variation of $k_z/h$ at fixed $k_z^+$ is realized through a change in Reynolds number -- the ratio of the outer and inner flow scales. This result suggests that for the determination of $RA/RA_0$, a mixed scaling should be considered, in which the roughness scale as well as both flow scales are present. Due to high computational cost, it is not possible in the present work to extend the range of Reynolds numbers in a way that a general scaling law can be extracted. Ideally experimental data following the principles established in this section (\ref{sec:results22}), but at an order of magnitude higher Reynolds numbers, should be used to complement the present data towards an accountable scaling law. 

\subsection{\label{sec:results3}Parametrization based on $k_s^+$}
\begin{figure}[h!]
	\begin{center}
		\includegraphics*[trim={0cm 0cm 1.5cm 1cm},clip,width=0.6\textwidth]{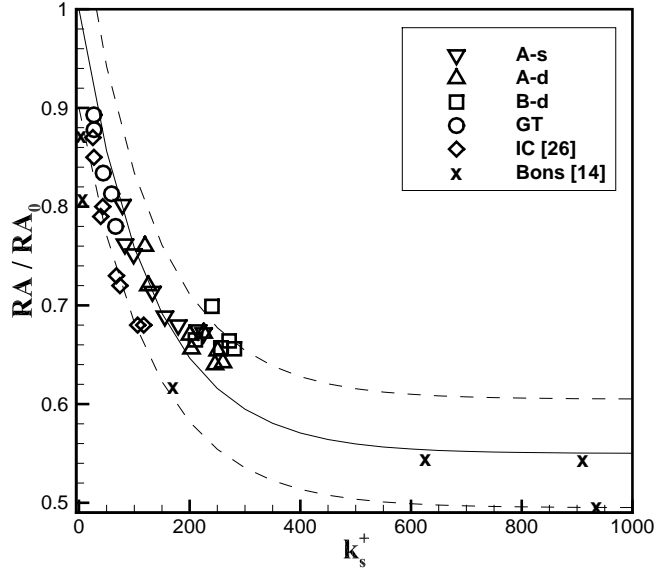}
		\caption{\label{fig:final} Variation of $RA/RA_0$ with inner scaled equivalent sand-grain roughness height $k_s^+$ for all simulations discussed in the present paper as well as experimental data from the boundary layer experiments of Bons \cite{Bons05}. Solid and dashed lines indicate Eq. \ref{eq:Forooghi} and its $\pm10\%$ error interval, respectively.}
	\end{center}
\end{figure}

In the previous sections, the effects of roughness morphology and its physical height on the Reynolds analogy were studied, and it was shown that both factors, as well as the scales of flow itself, can influence the Reynolds analogy factor ratio $RA/RA_0$. It is, however, of high practical value if one can find a general correlation in which different effects are to some degree contained. In a quest for a general yet simple correlation, we plot all the data points obtained from the 25 presently simulated cases and those from \cite{Forooghi18} against the inner scaled equivalent sand-grain roughness $k_s^+$ in figure \ref{fig:final}. Due to the limitation in Reynolds number faced in DNS, the highest value of $k_s^+$ in the simulations is smaller than 300. This shortcoming is, however, partly relieved using the experimental report published by Bons \cite{Bons05}, which provides a number of data points at higher values of $k_s^+$ up to 1000 (data reproduced from table 1 and figure 8 in \cite{Bons05}).

The collapse of the data points in figure \ref{fig:final} is much better than what can be obtained from any of the physical roughness parameters studied in the previous sections thanks to the fact that $k_s$ implicates the effects of both roughness morphology and height simultaneously. It should be noted that, the experiments in \cite{Bons05} are carried out for developing boundary layers subject to free stream turbulence at considerably higher Reynolds number ($Re_x=900,000$) than the present work. Despite the differences in the flow type and Reynolds number, the agreement between the results, in the overlapping part of the abscissa, is satisfactory. The present data points are somewhat above those reported by Bons, which is likely due to a combined effect of flow-type and Reynolds number and possibly variation of thermoplastic properties neglected in the present work.

A majority of the data points in figure \ref{fig:final} can be predicted within $\pm10\%$ error interval by the following correlation.
\begin{equation}
\frac{RA}{RA_0}=0.55+0.45e^{-k_s^+/130}
\label{eq:Forooghi}
\end{equation}
We follow \cite{Bons05} in adopting an exponential form of correlation that satisfies two constraints: $RA/RA_0=1$ at $k_s^+=0$ and $RA/RA_0$ tends to a constant value when $k_s^+$ is large enough. In our correlation, we opt for use of $k_s^+$ instead of the outer scaled $k_s$ employed by the above reference. Considering that the observed reduction in the Reynolds analogy factor of the rough walls is mainly a consequence of the departure from the state where viscous transport dominates at the wall, we believe that the ratio of roughness and viscous scales ($k_s^+$) is the most relevant factor in determining the Reynolds analogy factor over rough walls. One can also use the following analytical relation derived by Aupoix \cite{Aupoix15} to gain further insight into the problem.
\begin{equation}
\frac{RA}{RA_0}=\frac{1-\Delta U^+ \sqrt{c_{f,0}/2}}{1-RA_0 \Delta T^+ \sqrt{c_{f,0}/2}}
\label{eq:Aupoix}
\end{equation}
In Eq. \ref{eq:Aupoix}, $\Delta T^+$ is the logarithmic shift in the temperature profile or thermal analogous of the roughness function $\Delta U^+$, and $RA_0$ and $c_{f,0}$ stand for $RA$ and $C_f$ for a smooth wall at the same friction Reynolds number. The effect of roughness enters this relation through $\Delta T^+$ and $\Delta U^+$. Assuming an analogy between the momentum and thermal roughness functions in that they can be uniquely parametrized by $k_s^+$ (at constant Prandtdl number), the right hand side of Eq. \ref{eq:Aupoix} is only a function of $k_s^+$ except for the Reynolds number dependence of $c_{f,0}$. Considering that $\sqrt{c_{f,0}}$ varies only mildly with Reynolds number, it is reasonable to seek a correlation of $RA/RA_0$ in terms of $k_s^+$. 

Finally, Eq. \ref{eq:Forooghi} should not be seen as a perfect correlation accounting for all possible effects, but rather the best possible correlation using a single parameter. As shown in figure \ref{fig:final}, despite the many different geometries represented in this figure, and in particular, the large Reynolds number gap between the present data and those of Bons \cite{Bons05}, data points do not deviate considerably when correlated against $k_s$. The correlation would obviously benefit from further modifications using numerical and experimental data at different Reynolds numbers, roughness morphologies, and flow types.

\section{\label{sec:conclusion}Conclusions}

Direct Numerical Simulations were carried out in fully developed rough channels at friction Reynolds numbers up to 500 to investigate the effect of roughness on turbulent convection heat transfer. The major goal of this study was to understand how the Reynolds analogy factor $RA$ for a rough wall, normalized with the corresponding smooth-wall value $RA_0$, responds to a variation in the roughness morphology and the scaling of its height. A total of 25 simulations were carried out, in which the exact rough surface geometries were reproduced using an Immersed Boundary Method. Both artificial (random distribution of isolated roughness elements) and realistic roughness geometries were investigated. In the first part of the paper, the effect of roughness morphology at constant roughness height was studied. Here, the mean peak-to-valley height is used as the characteristic roughness height. In the second part of the paper, the effect of a variation in the characteristic roughness height -- in either inner or outer units -- while the roughness morphology remains unchanged was investigated. The most important findings are summarized below.

\begin{itemize}
\item[-] Reynolds analogy factor varies meaningfully with the investigated morphological properties, namely, mean slope, spacing and shape of the roughness elements. This variation is however moderate except for the roughness with very low slope or extremely sparse roughness, where the ratio $RA/RA_0$ abruptly increases and tends to unity. Based on the present data, a threshold of $\Lambda_s>50$ can be suggested for this abrupt increase, where $\Lambda_s$ is the density parameter of van Rij et al. \cite{vanRij02}.
\item[-] Frontal solidity and the density parameter $\Lambda_s$ were compared for their capability to correlate the effect of roughness morphology on $RA/RA_0$, and it was found that the latter has a relatively better performance.
\item[-] When the roughness morphology is unchanged, the ratio $RA/RA_0$ decreases with an increase in the roughness height in both inner and outer scales. The results point out the presence of a mixed scaling, which means that, in general case, both the inner and outer scales of flow as well as that of roughness should be considered in the determination of $RA$.
\item[-] It was shown that the ratio $RA/RA_0$ correlates relatively well with the inner-scaled equivalent sand roughness $k_s^+$ despite all different effects discussed in the paper. A correlation (Eq. \ref{eq:Forooghi}) were suggested based on the present DNS data and the experimental data from boundary layer flows at much higher Reynolds numbers reported by Bons \citep{Bons05}.
\end{itemize}

It is necessary to remind that the roughness geometries studied in the present paper belong to the k-type roughness or, roughly speaking, 3D random roughness. It is possible that the rules derived here do not work for fundamentally different types of geometries, for example 2D ribs or highly inhomogeneous roughness. 

As a final remark, although we attempted to study the problem at a wide range of conditions, further investigations of Reynolds analogy over rough walls will be highly beneficial in verifying and possibly modifying the present findings. In particular the Reynolds number limitation of DNS calls for supplementary experimental data at higher Reynolds numbers for the derivation of a universal scaling law. Further investigations of the problem for other flow types (e.g. boundary layers with conditions relevant to the flow over turbine blades) and roughness types (particularly realistic surfaces with low or negative skewness) will also be instrumental for establishing generalized rules.

\section*{Acknowledgements}
PF and BF acknowledge financial support by the German Research Foundation (DFG) under Collaborative Research Centre SFB/TRR150. The computations for this research were carried out on HPC cluster ForHLR1 at Karlsruhe Institute of Technology.

\end{document}